\RequirePackage{lineno}
\documentclass[prl,superscriptaddress,showpacs,amssymb,amsmath,amsfonts,aps]{revtex4}
\usepackage{graphicx}% Include figure files
\usepackage{epsfig}% Include figure files
\usepackage{dcolumn}% Align table columns on decimal point
\usepackage{bm}% bold math
\def\be{\begin{equation}}  
\def\ee{\end{equation}}  
\def\bea{\begin{eqnarray}}  
\def\eea{\end{eqnarray}}  
\def\tpI{$e p \rightarrow e \pi^+ n$ }

\def\tpIV{$e d \rightarrow e \pi^- p (p)$ }
\def\tpV{$e p \rightarrow e \pi^+ (n)$ }

\def\tpIn{$e p \rightarrow e \pi^+ n$}

\def\tpIVn{$e d \rightarrow e \pi^- p (p)$}
\def\tpVn{$e p \rightarrow e \pi^+ (n)$}

\def\tpVIII{$e d \rightarrow e \pi^- (pp)$ }

\def\tpVIIIn{$e d \rightarrow e \pi^- (2p)$}

\def\mxepi{$M_x^{e\pi}$ }

\def\mxep{$M_x^{eN} $}

\def\cthcm{$\cos(\theta^*)$}
\def\cthcmsp{$\cos(\theta^*)$ }
\def\cthcmn{\cos(\theta^*)}
\def\phicm{$\phi^*$}
\def\phicmsp{$\phi^*$ }

\def\>{\hskip .1in }

\begin{document}
%\linenumbers

%\preprint{CLAS-ANALYSIS-NOTE}
\title{Target and Beam-Target Spin Asymmetries in Exclusive 
$\pi^+$ and $\pi^-$ Electroproduction  
with 1.6 to 5.7 GeV Electrons}

\newcommand*{\ANL}{Argonne National Laboratory, Argonne, Illinois 60439}
\newcommand*{\ANLindex}{1}
\affiliation{\ANL}
\newcommand*{\CSUDH}{California State University, Dominguez Hills, Carson, CA 90747}
\newcommand*{\CSUDHindex}{2}
\affiliation{\CSUDH}
\newcommand*{\CANISIUS}{Canisius College, Buffalo, NY}
\newcommand*{\CANISIUSindex}{3}
\affiliation{\CANISIUS}
\newcommand*{\CMU}{Carnegie Mellon University, Pittsburgh, Pennsylvania 15213}
\newcommand*{\CMUindex}{4}
\affiliation{\CMU}
\newcommand*{\CUA}{Catholic University of America, Washington, D.C. 20064}
\newcommand*{\CUAindex}{5}
\affiliation{\CUA}
\newcommand*{\SACLAY}{CEA, Centre de Saclay, Irfu/Service de Physique Nucl\'eaire, 91191 Gif-sur-Yvette, France}
\newcommand*{\SACLAYindex}{6}
\affiliation{\SACLAY}
\newcommand*{\CNU}{Christopher Newport University, Newport News, Virginia 23606}
\newcommand*{\CNUindex}{7}
\affiliation{\CNU}
\newcommand*{\UCONN}{University of Connecticut, Storrs, Connecticut 06269}
\newcommand*{\UCONNindex}{8}
\affiliation{\UCONN}
\newcommand*{\FU}{Fairfield University, Fairfield CT 06824}
\newcommand*{\FUindex}{9}
\affiliation{\FU}
\newcommand*{\FIU}{Florida International University, Miami, Florida 33199}
\newcommand*{\FIUindex}{10}
\affiliation{\FIU}
\newcommand*{\FSU}{Florida State University, Tallahassee, Florida 32306}
\newcommand*{\FSUindex}{11}
\affiliation{\FSU}
\newcommand*{\Genova}{Universit$\grave{a}$ di Genova, 16146 Genova, Italy}
\newcommand*{\Genovaindex}{12}
\affiliation{\Genova}
\newcommand*{\GWUI}{The George Washington University, Washington, DC 20052}
\newcommand*{\GWUIindex}{13}
\affiliation{\GWUI}
\newcommand*{\ISU}{Idaho State University, Pocatello, Idaho 83209}
\newcommand*{\ISUindex}{14}
\affiliation{\ISU}
\newcommand*{\INFNFE}{INFN, Sezione di Ferrara, 44100 Ferrara, Italy}
\newcommand*{\INFNFEindex}{15}
\affiliation{\INFNFE}
\newcommand*{\INFNFR}{INFN, Laboratori Nazionali di Frascati, 00044 Frascati, Italy}
\newcommand*{\INFNFRindex}{16}
\affiliation{\INFNFR}
\newcommand*{\INFNGE}{INFN, Sezione di Genova, 16146 Genova, Italy}
\newcommand*{\INFNGEindex}{17}
\affiliation{\INFNGE}
\newcommand*{\INFNRO}{INFN, Sezione di Roma Tor Vergata, 00133 Rome, Italy}
\newcommand*{\INFNROindex}{18}
\affiliation{\INFNRO}
\newcommand*{\INFNTUR}{INFN, Sezione di Torino, 10125 Torino, Italy}
\newcommand*{\INFNTURindex}{19}
\affiliation{\INFNTUR}
\newcommand*{\ORSAY}{Institut de Physique Nucl\'eaire, CNRS/IN2P3 and Universit\'e Paris Sud, Orsay, France}
\newcommand*{\ORSAYindex}{20}
\affiliation{\ORSAY}
\newcommand*{\ITEP}{Institute of Theoretical and Experimental Physics, Moscow, 117259, Russia}
\newcommand*{\ITEPindex}{21}
\affiliation{\ITEP}
\newcommand*{\JMU}{James Madison University, Harrisonburg, Virginia 22807}
\newcommand*{\JMUindex}{22}
\affiliation{\JMU}
\newcommand*{\KNU}{Kyungpook National University, Daegu 41566, Republic of Korea}
\newcommand*{\KNUindex}{23}
\affiliation{\KNU}
\newcommand*{\MISS}{Mississippi State University, Mississippi State, MS 39762-5167}
\newcommand*{\MISSindex}{24}
\affiliation{\MISS}
\newcommand*{\UNH}{University of New Hampshire, Durham, New Hampshire 03824-3568}
\newcommand*{\UNHindex}{25}
\affiliation{\UNH}
\newcommand*{\NSU}{Norfolk State University, Norfolk, Virginia 23504}
\newcommand*{\NSUindex}{26}
\affiliation{\NSU}
\newcommand*{\OHIOU}{Ohio University, Athens, Ohio  45701}
\newcommand*{\OHIOUindex}{27}
\affiliation{\OHIOU}
\newcommand*{\ODU}{Old Dominion University, Norfolk, Virginia 23529}
\newcommand*{\ODUindex}{28}
\affiliation{\ODU}
\newcommand*{\URICH}{University of Richmond, Richmond, Virginia 23173}
\newcommand*{\URICHindex}{29}
\affiliation{\URICH}
\newcommand*{\ROMAII}{Universita' di Roma Tor Vergata, 00133 Rome Italy}
\newcommand*{\ROMAIIindex}{30}
\affiliation{\ROMAII}
\newcommand*{\MSU}{Skobeltsyn Institute of Nuclear Physics, Lomonosov Moscow State University, 119234 Moscow, Russia}
\newcommand*{\MSUindex}{31}
\affiliation{\MSU}
\newcommand*{\SCAROLINA}{University of South Carolina, Columbia, South Carolina 29208}
\newcommand*{\SCAROLINAindex}{32}
\affiliation{\SCAROLINA}
\newcommand*{\TEMPLE}{Temple University,  Philadelphia, PA 19122 }
\newcommand*{\TEMPLEindex}{33}
\affiliation{\TEMPLE}
\newcommand*{\JLAB}{Thomas Jefferson National Accelerator Facility, Newport News, Virginia 23606}
\newcommand*{\JLABindex}{34}
\affiliation{\JLAB}
\newcommand*{\UTFSM}{Universidad T\'{e}cnica Federico Santa Mar\'{i}a, Casilla 110-V Valpara\'{i}so, Chile}
\newcommand*{\UTFSMindex}{35}
\affiliation{\UTFSM}
\newcommand*{\EDINBURGH}{Edinburgh University, Edinburgh EH9 3JZ, United Kingdom}
\newcommand*{\EDINBURGHindex}{36}
\affiliation{\EDINBURGH}
\newcommand*{\GLASGOW}{University of Glasgow, Glasgow G12 8QQ, United Kingdom}
\newcommand*{\GLASGOWindex}{37}
\affiliation{\GLASGOW}
\newcommand*{\VT}{Virginia Tech, Blacksburg, Virginia   24061-0435}
\newcommand*{\VTindex}{38}
\affiliation{\VT}
\newcommand*{\VIRGINIA}{University of Virginia, Charlottesville, Virginia 22901}
\newcommand*{\VIRGINIAindex}{39}
\affiliation{\VIRGINIA}
\newcommand*{\WM}{College of William and Mary, Williamsburg, Virginia 23187-8795}
\newcommand*{\WMindex}{40}
\affiliation{\WM}
\newcommand*{\YEREVAN}{Yerevan Physics Institute, 375036 Yerevan, Armenia}
\newcommand*{\YEREVANindex}{41}
\affiliation{\YEREVAN}

\newcommand*{\NOWJLAB}{Thomas Jefferson National Accelerator Facility, Newport News, Virginia 23606}
\newcommand*{\NOWUK}{University of Kentucky, Lexington, KY 40506}
\newcommand*{\NOWORSAY}{Institut de Physique Nucl\'eaire, CNRS/IN2P3 and Universit\'e Paris Sud, Orsay, France}
\newcommand*{\NOWGLASGOW}{University of Glasgow, Glasgow G12 8QQ, United Kingdom}
\newcommand*{\NOWINFNGE}{INFN, Sezione di Genova, 16146 Genova, Italy}
\newcommand*{\NOWSSI}{Spectral Sciences Inc., Burlington, MA 01803}

% Ralph Minehart said ``no'' to lead authorship
% got comments from Rob, Sebastian, Josh, and Yelena
% author check: added Charles, Fanchini, Jo, Procureur, added INRNF to Rossi, 
\author{P.E.~Bosted}
     \email{bosted@jlab.org}
\affiliation{\WM}
\author {A.S.~Biselli} 
\affiliation{\FU}
\author{S.~Careccia}
\affiliation{\ODU}
\author{G.~Dodge}
\affiliation{\ODU}
\author {R.~Fersch} 
\affiliation{\CNU}
\author{N. Guler}
\altaffiliation[Current address: ]{\NOWSSI}
\affiliation{\ODU}
\author {S.E.~Kuhn} 
\affiliation{\ODU}
\author{J.~Pierce}
\affiliation{\JLAB}
\author {Y.~Prok} 
\affiliation{\ODU}
\author{X. Zheng}
\affiliation{\VIRGINIA}
% ...

\collaboration{The CLAS Collaboration}
\noaffiliation

\date{\today}


\begin{abstract}
Beam-target double spin asymmetries and target single-spin
asymmetries in exclusive  $\pi^+$
and quasi-exclusive $\pi^-$  electroproduction
were obtained from scattering of 1.6 to 5.7 GeV longitudinally polarized
electrons from longitudinally polarized protons (for $\pi^+$)
and deuterons (for $\pi^-$)
using the CEBAF Large Acceptance Spectrometer 
(CLAS) at Jefferson Lab. 
The kinematic range covered is $1.1<W<2.6$ GeV and $0.05<Q^2<5$
GeV$^2$, with good angular coverage in the forward
hemisphere. The asymmetry results were divided into 
approximately 40,000 kinematic bins 
for $\pi^+$ from free protons and 15,000 bins for $\pi^-$
production from bound nucleons in the deuteron. The present 
results are found to be in reasonable agreement
with fits to previous world data for $W<1.7$ GeV and 
$Q^2<0.5$ GeV$^2$, with discrepancies increasing
at higher values of $Q^2$, especially for $W>1.5$ GeV.
Very large target-spin asymmetries are observed for
$W>1.6$ GeV. When combined with cross section measurements,
the present results can provide powerful constraints
on nucleon resonance amplitudes at moderate and large
values of $Q^2$, for resonances with masses as high as 2.3 GeV.
\end{abstract}
\maketitle

%\tableofcontents

\section{Introduction}

Exclusive electroproduction of pseudo-scalar mesons is
a process that is sensitive to the detailed internal
structure of the nucleon. The process is particularly
sensitive to contributions from individual 
nucleon resonance states. Photoproduction and very
low $Q^2$ electroproduction continue to provide
insight into the static properties of the resonances,
such as mass, width, parity, spin, and decay branching
ratios. Larger values of $Q^2$ are needed to study
transition form factors, and also reveal the existence
of resonances that are suppressed in photoproduction.
Initial large-$Q^2$ measurements of spin-averaged 
cross sections for exclusive $\pi^+$ electroproduction from 
Cornell~\cite{bebek76,bebek78} had limited statistical
accuracy. Recent measurements from Jefferson Lab
($\rm{JLab}$)~\cite{Horn09,HPBlok,XQian,ParkA,ParkB,ParkC} 
have greatly improved the situation. 
A relatively limited
data set exists for exclusive $\pi^-$ 
electroproduction (including 
Refs.~\cite{bebek76,bebek78,morris,Fosteretc}).

The use of polarized nucleon targets and polarized
electron beams
is particularly useful in distinguishing between
resonances of different spin, isospin, and parity, because
all single-spin asymmetries vanish in the absence of 
interference terms. Beam asymmetries at large $Q^2$
were published from JLab for $W<1.7$ GeV \cite{ParkA}
and are also the subject of an early 
investigation for $W>2$ GeV~\cite{Avakian}. 
Beam-target asymmetries for positive
pions were reported from a pilot experiment
at Jefferson Lab~\cite{daVita}. 

The present experiment provides the first
body of high-statistical precision target and beam-target
asymmetries spanning a wide range of $Q^2$ and $W$,
for both $\pi^+$ and $\pi^-$  electroproduction.
The $\pi^+$ results are complementary to  results
from two other Jefferson Lab experiments, named eg4~\cite{eg4}
and eg1-dvcs~\cite{eg1dvcs}, focused on the low $Q^2$ and
high $Q^2$ regions, respectively.

After a summary of the formalism, details of the experimental
setup, analysis, and results are presented in the following
sections.

\section{Formalism}
\label{sec:formalism}
Because both
the beam and the target were longitudinally polarized,
we could, in principle, extract three spin asymmetries,
defined by:
\be
\sigma = \sigma_0 (1 + P_B A_{LU} + P_T A_{UL} + P_BP_T A_{LL}),
\ee
where $P_B$ and $P_T$ are the longitudinal beam and target polarizations,
respectively, $\sigma_0$ is the spin-averaged
cross section, and $A_{LU}$, $A_{UL}$, and $A_{LL}$ are
the beam, target, and beam-target asymmetries, respectively.
The cross sections and asymmetries are all
functions of five independent
variables. For this analysis, 
the variables $(W, Q^2, \cos(\theta^*), \phi^*, \epsilon)$
are used, where $\theta^*,\phi^*$ are the center-of-mass
decay angles of the final state with invariant mass $W$ into a meson and 
a nucleon, $Q^2$ is the squared virtual photon four-momentum,
and $\epsilon$ is the virtual photon polarization. The bins
in $\epsilon$ have a one-to-one correlation with the
different beam energies of the experiment.
 We use the convention that the center-of-mass final state
decay polar angle $\theta^*=0$ degrees
corresponds to a forward-going meson. The definition 
of $\phi^*$ is the opening angle between
$(\vec q \times \vec e)$ and $(\vec q \times \vec p_\pi)$,
where $\vec e$ is the incident electron momentum,
$\vec q$ is the momentum transfer to the scattered
electron, and $\vec p_\pi$ is the detected pion momentum.

Following the conventions of the MAID group~\cite{maid},
the beam and target asymmetries can be expressed as:
\be 
A_{LL}=-\sigma_{ez} / \sigma_0
\ee
\be 
A_{UL}= \sigma_{z} / \sigma_0,
\ee
where
$$ \sigma_{ez}  \,=  \,\sqrt{2 \epsilon (1-\epsilon)} \,
       [P_x \sigma_{TL'x}  \cos(\phi^*)  \, + \, 
        P_y \sigma_{TL'y}  \sin(\phi^*)  \, +  \,
        P_z \sigma_{TL'z}  \cos(\phi^*)]  \, + $$
$$       \sqrt{1-\epsilon^2}   \,
        (P_x  \sigma_{TT'x}  \, + \, 
         P_z  \sigma_{TT'z}),
$$
$$ \sigma_z  \, =  \, \sqrt{2\epsilon (1+\epsilon)}  \, 
       (P_x  \sigma_{TLx} \sin(\phi^*)  \, +  \,
        P_y  \sigma_{TLy} \cos(\phi^*)   \, +  \,
        P_z  \sigma_{TLz} \sin(\phi^*)  ) \, + $$
$$       \epsilon   \,
       (P_x  \sigma_{TTx} \sin(2\phi^*)  \, + \,
        P_y  \sigma_{TTy} \cos(2\phi^*)  \, + \,
        P_z  \sigma_{TTz} \sin(2\phi^*)  ) \, + \,
       P_y  (\sigma_{Ty} \, + \, \epsilon  \sigma_{Ly}),
$$
\noindent and
$$      \sigma_0 = \sigma_{T} + \, \epsilon  \sigma_{L} + 
          \, \sqrt{2\epsilon(1+\epsilon)} \, 
          \cos(\phi^*)  \sigma_{TL} \, + \, 
          \epsilon\cos(2\phi^*) \sigma_{TT}, 
$$
where the direction cosines are defined as $P_z=\cos(\theta_q)$, 
$P_y=-\sin(\theta_q)\sin(\phi^*)$, and
$P_x= \sin(\theta_q)\cos(\phi^*)$, and the virtual photon polarization as
$$\epsilon = 1/[1 + 2 (1 + \nu^2 / Q^2) \tan^2(\theta_e)],$$ 
where $\nu$ is the virtual photon energy.
The angles  $\theta_e$ and $\theta_q$ are relative to
the beam line direction for the scattered electron and
the momentum transfer, respectively. 
The cross sections 
$\sigma_{L}$, $\sigma_{T}$, $\sigma_{TL}$,
$\sigma_{TT}$, $\sigma_{TL'}$, and $\sigma_{TT'}$ are functions
of the three variables $W$, $Q^2$, and $\theta^*$.
%, and
%can be evaluated from the resonance helicity amplitudes
%and the non-resonant Born diagrams. 

In the case of $\pi^-$ electroproduction from polarized
deuterons, the above relations do not account for modifications
from the proper treatment of the 
deuteron wave-function (including the D-state in particular) 
as well as final state interactions (such as charge-exchange
reactions). These effects should be taken into account when
interpreting the asymmetries presented in this paper in terms
of reduced cross sections.

\section{Experiment}

The ``eg1b'' experiment used 
1.6 to 5.7 GeV longitudinally polarized electrons 
from CEBAF at Jefferson Lab impinging
on a 0.02 radiation length longitudinally polarized 
solid ammonia target immersed in liquid helium 
\cite{Keith}. The target polarization
direction is along the incident electron direction,
{\it not} the direction of the momentum transfer vector,
resulting in non-zero values of $P_x$ and $P_y$.
Scattered electrons and charged pions
were detected in the CEBAF Large Acceptance 
Spectrometer (CLAS)~\cite{CLAS}.
The typical beam current was a few nA.
The beam polarization, as periodically measured using
M\o ller scattering in an upstream polarimeter, averaged
70\%.

About 30\% of the running time was on polarized
protons ($^{15}$NH$_3$ target), 50\% on polarized deuterons
($^{15}$ND$_3$ target), 13\% on a reference unpolarized
carbon target, and 2\% on an empty cell (essentially a pure
helium target). The ammonia targets used the $^{15}$N isotope
to simplify polarized nitrogen corrections.  
The 1.5-cm-diameter cups typically contained
0.7 g/cm$^2$ of material immersed in a 2-cm-long
liquid-helium bath.  The aerial densities of the
target materials are listed in Table~\ref{tab:target}.
In this table, the thin Kapton foils that hold the ammonia
beads have been merged with the aluminum beam windows that 
contain the helium.
The composition depends on two parameters, $L$ and $l_A$,
whose values are listed in Table~\ref{tab:target}.

\begin{table}[hbt]
\begin{tabular}{cccccc}
Target & NH$_3$ & ND$_3$ & He & Al & C \\
\hline
NH$_3$ & $0.917 l_A $ &           - & $0.145(L-l_A)$   & 0.09 & - \\  
ND$_3$ &     -       & $1.056 l_A $ & $0.145(L-l_A)$   & 0.09 & - \\ 
Carbon &    -        &           - & $0.145(L-0.23)$ & 0.09 & 0.499 \\ 
\hline
\end{tabular}
\caption{Composition of the three targets used in this
analysis, in units of g/cm$^2$ as a function of the total length of the target
$L$ (range 1.8 to 2.2 cm) 
and effective ammonia length $l_A$ (range 0.53 to 0.73 cm).
}
\label{tab:target}
\end{table}

To reduce the rate of depolarization of the target 
from radiation damage, 
 the sub-millimeter-diameter 
beam was uniformly rastered over
the 1.5-cm-diameter front face of the target every few seconds. The
beam position, averaged over a few minutes or longer,
was kept stable at the 0.1 mm level, using feedback from
a set of beam position monitors. 
A split solenoid superconducting magnet provided a highly
uniform 5 T magnetic field surrounding the target 
($\delta B/B \approx 10^{-5}$).

Particles were detected in CLAS for  
polar angles from 8 to 48 degrees. CLAS 
comprises six azimuthally symmetric detector
arrays embedded in a toroidal magnetic field.
Charged particle momenta and scattering angles
were measured with the drift chamber 
tracking system. The momentum resolution ranged
from about 0.5\% at 0.5 GeV to over 2\% at 6 GeV. The 
resolution in polar angles was about 1 mrad, while the 
azimuthal angle resolution was typically 4 mrad.
Electrons were separated from a significantly 
larger flux of charged pions using segmented
gas Cherenkov detectors (CC, pion threshold 2.6 GeV)
and a sampling electromagnetic calorimeter (EC). 
A layer of time-of-flight scintillator counters (SC) between the
CC and EC was used for hadron identification.
In order to not overwhelm the 
data acquisition system, the hardware
trigger system was designed to have high efficiency
for events with a scattered electron with an energy
greater than 0.3 GeV, while rejecting other events.
The hardware Cherenkov
and calorimeter thresholds were adjusted to give
a trigger rate of about 3000 Hz, with a dead time
of about 10\%. The distance from the target to CLAS center was fixed
at about -55 cm for the entire run.

The data taking took place in late 2000 and early 2001. 
The data set
is divided into several Parts, each with a different
beam energy (2p5 for 2.5 GeV, etc.) and
specific  CLAS torus polarity (``i'' or ``o'' for electron
bending inward or outward in the torus). 
The field strength was three quarters of standard full strength  
(corresponding to 3000 A) for those Parts with beam energy 
above 4 GeV, and half of the standard value for the other Parts. 
A summary is given in Table~\ref{tab:Parts}.
Both the $^{15}$NH$_3$ (proton) and $^{15}$ND$_3$ (deuteron)
targets were used
for all Parts except Part 2p5i, which only had the deuteron target,
and Part 2p2i, which only had the proton target.
Part 1p6o was not included in the final analysis, because
data were taken only with the positive target polarization
direction (for both NH$_3$
and ND$_3$), and both directions are needed to form target spin
asymmetries. The relatively short Part 5p74o was not used because
of corruption of the data taken with the carbon target
(needed for luminosity normalization).
 Within each Part used, some short running
periods were removed due to problems with beam quality,
target polarization, or detector performance.

\begin{table}[hbt]
\begin{tabular}{lcrcccc}
Run period   & \> beam energy  & \> $I$ torus &
\> $P_BP_T$(p) & \> $P_BP_T$(d)& $R_{A>2}^p$ & $R_{A>2}^d$ \\
\hline
% order is my code 1,2,4,5,6,7,8,9,10,12,13,14
% (ie skipped 3 and 11)
Part 1p6i  & 1.603 GeV &  1500 A & 0.55 & 0.21 & 0.86 & 0.99\\
(Part 1p6o)  & 1.603 GeV & -1500 A & -  & -  & - & - \\
Part 1p7o  & 1.721 GeV & -1500 A & 0.58 & 0.21  & 0.81 & 0.99\\
Part 2p2i  & 2.285 GeV &  1500 A & 0.50 &  -    & 0.86 &  -  \\
Part 2p5i  & 2.559 GeV &  1500 A &  -   & 0.21  &  -   & 0.99\\
Part 2p5o  & 2.559 GeV & -1500 A & 0.61 & 0.25  & 0.86 & 1.01\\
Part 4p2i  & 4.236 GeV &  2250 A & 0.54 & 0.18  & 0.85 & 0.99\\
Part 4p2o  & 4.236 GeV & -2250 A & 0.55 & 0.18  & 0.88 & 1.01\\
Part 5p6i  & 5.612 GeV &  2250 A & 0.50 & 0.20  & 0.815 & 0.99\\
Part 5p72i & 5.722 GeV &  2250 A & 0.50 & 0.20  & 0.815 & 0.99\\
Part 5p72o & 5.722 GeV & -2250 A & 0.50 & 0.19  & 0.83 & 0.99\\
(Part 5p74o) & 5.740 GeV & -2250 A & 0.50 & 0.19  & - & -\\
\hline
\end{tabular}
\caption{Run period names, electron beam energy, and CLAS torus
current of the different parts of the experiment analyzed.
Also listed are the product of the absolute value of 
beam and target polarization
for the polarized proton and deuteron 
runs (see Sec.~\ref{sec:pbpt}). The last
two columns list the ratios of bound protons in the
NH$_3$ and carbon targets ($R_{A>2}^p$) and bound neutrons in
the ND$_3$ and carbon targets ($R_{A>2}^d$) (see Sec.~\ref{sec:df}
for full details).
}
\label{tab:Parts}
\end{table}

One of the primary goals of the eg1b experiment 
was the measurement of spin structure functions through
inclusive electron scattering, with results reported in
Refs.~\cite{prok,vipuli,duality,inclp,incld}. Many experimental details
can be found in these publications. Results
for the other primary goal, which is the determination of
charged pion electroproduction spin asymmetries, 
are the subject of the present paper and two
Ph.D theses~\cite{pierce,sharon}. Results have also
been published for neutral pion electroproduction spin 
asymmetries for the lowest beam energy of the present
experiment~\cite{bissellieg1b}.

\section{Analysis}

\subsection{Data Processing}
A subset of the data was used to calibrate the response
of all of the CLAS detectors. The instruments that
measured beam position and current were calibrated.
The alignment of the detectors, as well as the target
magnet, were also determined.

The raw data were passed through a standard CLAS analysis
package that transformed raw timing and pulse-height
signals into a set of ``particles'' for each trigger event.
Direction cosines at the target were determined from
the drift chambers for charged particles, and from
the hit positions in the EC in the
case of neutral particles. The momenta of charged
particles were determined from the drift chamber tracks,
while the energy of neutrals was determined from the
EC. Charged-particle tracks were associated
with the corresponding CC signals, EC
energy deposition, and timing from the SC using geometrical
matching. Additional details can be found in the two archival papers 
describing the eg1b inclusive analysis~\cite{inclp,incld}.

A subset of the recorded events were subsequently written to 
skimmed data files for further processing. These data files
only contained events that had a reasonable chance of
passing the event selection cuts of the present analysis.
 
\subsection{Particle Identification}
In the present analysis, we are interested
in two reactions, \tpI and \tpIVn. For each reaction, we
analyzed two distinct topologies, which were later combined. 
The four topologies are listed in Table~\ref{tab:particles},
along with the particles that must be identified in each case.
The analysis of $\pi^-$ electroproduction made the assumption
that the interaction took place on a neutron and that the
spectator proton was ``invisible'': therefore both 
\tpI and \tpIV are referred to as ``fully exclusive'' topologies,
and \tpV and \tpVIII are referred to as ``one-missing particle''
topologies in the remainder of this article. 

\begin{table}[hbt]
\begin{tabular}{ll}
topology & final state particles \\
\hline 
\tpI    & electron, $\pi^+$, neutron \\
\tpIV   & electron, $\pi^-$, proton  \\
\tpV    & electron, $\pi^+$  \\
\tpVIII & electron, $\pi^-$          \\
\
\end{tabular}
\caption{Particles to be identified for each of the
topologies of this analysis.
}
\label{tab:particles}
\end{table}

The fully exclusive and one-missing-particle topologies
are distinct: in making the skim files, an event was put
in the fully exclusive topology if a detected nucleon
passed loose exclusivity cuts, else it was stored in
the non-exclusive topology. If a fully exclusive event
did not pass the slightly stricter cuts at the second 
level of processing, it was not moved over to the 
non-exclusive topology. Rather, the event was discarded
completely, because such events predominantly originate
from the nitrogen in the target.

\subsubsection{Electron Identification}
Electrons were identified by requiring a signal of
at least 2.5  photo-electrons in the CC, 
at least 67\% of the electron energy to be
deposited in the EC (front and back layers combined), 
and at least 6\% of the electron energy to be deposited
in the front layer of the EC. These cuts were needed to separate
electrons from pions, which would otherwise overwhelm the electron
sample at the higher beam energies of this experiment. 
The track vertex position was required to be 
reconstructed within 3 cm of the nominal target center to 
remove backgrounds from the target chamber windows and heat
shield foils.  
An additional cut to reduce pion contamination 
required that the track position in the SC 
be matched to the position in the CC.

\subsubsection{Charged Pion and Proton Identification}
Charged hadrons were identified by requiring that
the time-of-arrival at the scintillator
counters be within 0.8 ns of that predicted from
the time-of-arrival of the electron in the event.
It was further required that charged pions and protons
do not produce a significant signal in the CC
(i.e. less than one photo-electron).
A vertex cut of $\pm 3$ cm was also required.
Finally, particles produced at polar angles greater than 48
degrees in the lab frame were rejected because they passed
through thick materials, causing significant energy loss and
multiple scattering. 
 
\subsubsection{Neutron Identification}
Neutrons were identified by requiring the absence of a drift chamber 
track and a 
time-of-arrival at the EC corresponding to $\beta<0.95$ to 
separate neutrons from photons. A further cut required an energy
deposit of at least 0.3 GeV in the EC, to separate neutrons from
low energy photons originating from an out-of-time interaction.
The direction cosines of the neutron were determined
from the EC hit coordinates. As discussed later, neutrons were only
used to obtain a better dilution factor in 
exclusive $\pi^+$ production, so the cases where
the neutron was not identified simply moved
events from \tpI to \tpVn. 
The neutron momentum
could not be determined from time-of-flight with
sufficient accuracy to be useful.

\subsection{Exclusivity Kinematic Cuts}

For each of the four topologies, kinematic cuts
were placed to improve the signal-to-noise ratio. The value
of kinematic cuts is two-fold. First,
most of the kinematic quantities have a wider
distribution for bound nucleons (in target materials
with atomic number $A>2$) than for free protons (or almost free
neutrons in the deuteron). Kinematic cuts therefore
reduce the dilution of the signal of interest
(scattering from polarized free protons or quasi-free 
neutrons), compared to the
background from unpolarized nucleons in materials
with $A>2$. Furthermore, kinematic cuts are needed to
isolate single meson production from multi-meson
production. 

Many different kinematic cuts were found to be useful.
All topologies used a cut on electron-pion missing
mass. The topologies \tpI and \tpIV 
had additional cuts on the angles of the recoil nucleon.
Topology \tpIV had additional cuts on the electron-proton
missing mass and the electron-pion-proton missing energy.
Details on all of these cuts are given in the sections below.

\subsubsection{Electron-Pion Missing Mass Cuts}
In all of the topologies studied,
the electron-pion missing mass \mxepi
should be equal to the nucleon mass $M$.
In general, one would like the upper cut on \mxepi
to be well below  $M+m_\pi=1.08$ GeV, to avoid
contributions from multi-pion production, where
$m_\pi$ is the pion mass. Placing
tighter cuts helps to reduce the nuclear background.

\begin{figure}[hbt]
%\centerline{\includegraphics[height=7.0in,angle=90]{exclwwwprc1.pdf}}
\centerline{\includegraphics[height=7.0in,angle=90]{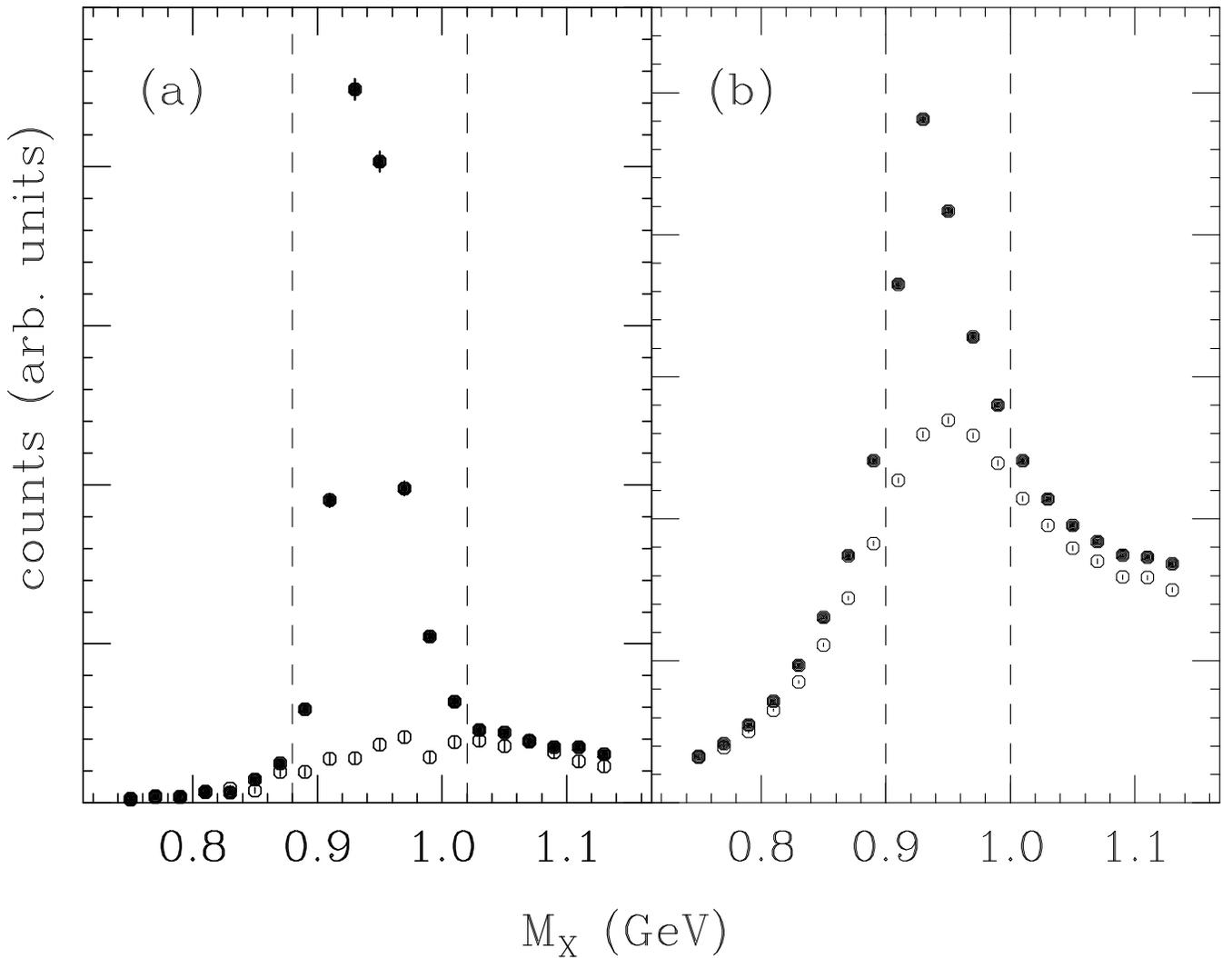}}
%\scalebox{1.0}[1.0]{\includegraphics{exclwwwprc1.ps}}

\caption{Electron-pion missing mass spectra
from Part 4p2o for the topology \tpI (a)
and topology \tpV (b).
Counts from the ammonia NH$_3$ target are shown
as the solid circles and counts from the carbon target
(scaled by the ratio of integrated luminosities on 
bound nucleons) are shown as the open circles. 
}
\label{fig:www1}
\end{figure}

The spectra for \mxepi for topologies
\tpI  and \tpV are shown in Fig.~\ref{fig:www1}. 
For \tpVn, the missing mass
was calculated assuming quasi-free production from a 
neutron in the deuteron. 
The spectra are from Part 4p2o. 
The other cuts used for the \tpI
topology have been applied (no other cuts were used
for \tpVn).
The solid circles correspond to counts from the ammonia
target, while the open circles correspond to counts
from the carbon target, scaled by the ratio of
luminosities on $A>2$ nucleons. 
A clear peak is visible near the nucleon mass (0.94 GeV) 
from the ammonia target, with a smaller and much 
wider distribution from the carbon target. The wings of the ammonia
distributions match well to the scaled carbon spectra 
on the low-mass side of the peaks, demonstrating that differences
between nitrogen and carbon (and to a much smaller extent helium)
due to Final State Interaction (FSI), Fermi motion, and other
possible nuclear effects are relatively minor. On
the high side of the peaks, the ammonia rates are
higher, due to the radiative tail of the single-pion
production. For the fully exclusive topology, the
nuclear background is very small, while for the
non-exclusive topology, the typical background is about
half the signal size.

\begin{figure}[hbt]
%\centerline{\includegraphics[height=7.0in,angle=90]{exclwwwprc4.pdf}}
\centerline{\includegraphics[height=7.0in,angle=90]{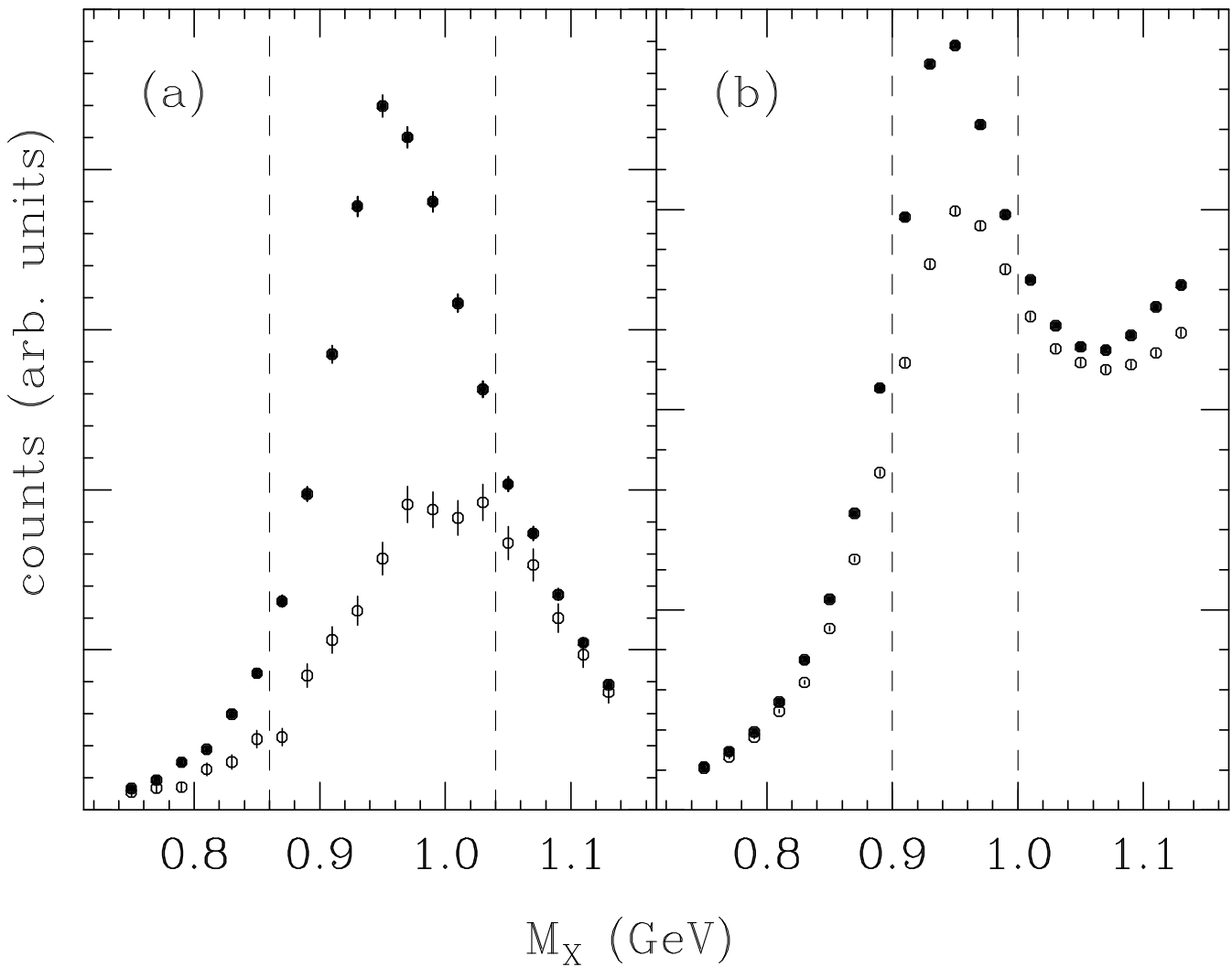}}
\caption{Electron-pion missing mass spectra
from Part 4p2o for the topology \tpIV (a)
and topology \tpVIII (b).
Counts from the ammonia ND$_3$ target are shown
as the solid circles and counts from the carbon target
(scaled by the ratio of integrated luminosities on 
bound nucleons) are shown as the open circles. 
}
\label{fig:www4}
\end{figure}

The spectra for \mxepi are shown for the two $\pi^-$ topologies
in Fig.~\ref{fig:www4}a,b.
The peaks from the ND$_3$ target are wider than the
corresponding peaks in the 
positive topologies due to the Fermi motion
of neutrons in deuterium (which is about four times
smaller than in $A>2$ nuclei). This results in a larger
nuclear background for the negative pion topologies
than for the positive pion topologies. 

The dashed vertical lines show the cuts used to minimize
the final asymmetry uncertainties. The same cuts were used for
all beam energies. The cut values are listed in Table~\ref{tab:www}.

\begin{table}[hbt]
\begin{tabular}{ll}
topology & cut range \\
\hline 
\tpI    & $0.88<M_x^{e\pi}<1.02$ GeV \\
\tpIV   & $0.86<M_x^{e\pi}<1.04$ GeV \\
\tpV    & $0.90<M_x^{e\pi}<1.00$ GeV \\
\tpVIII & $0.90<M_x^{e\pi}<1.00$ GeV \\
\end{tabular}
\caption{Lower and upper limits of the cuts used
on \mxepi for each of the relevant topologies.
}
\label{tab:www}
\end{table}

\subsubsection{Electron-Proton Missing Mass Cuts}
\begin{figure}[hbt]
\centerline{\includegraphics[height=7.0in,angle=90]{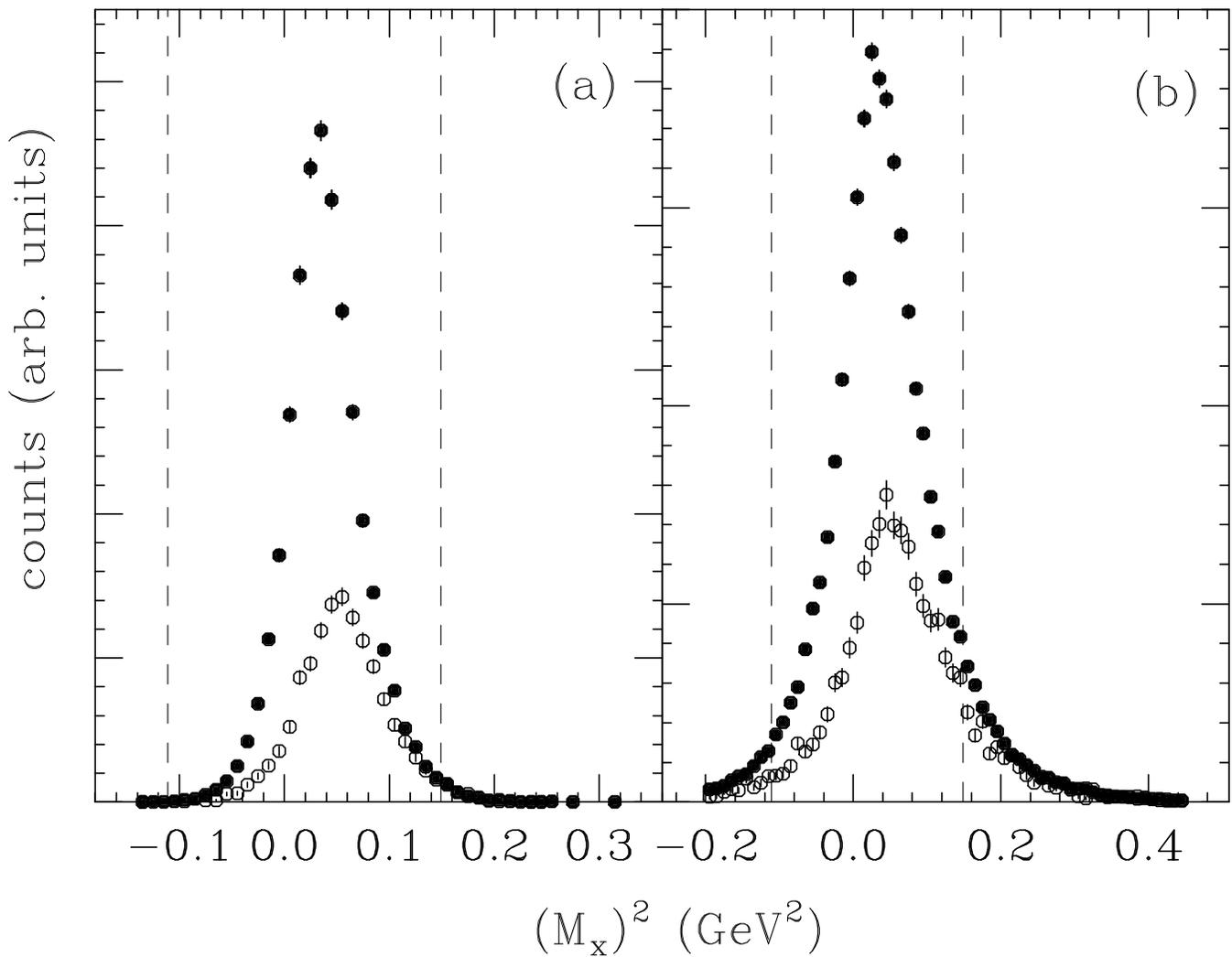}}
\caption{Squared electron-proton missing mass spectra for the
topology \tpIV for Part 1p6i (a) and Part 4p2o
(b).
Counts from the ND$_3$ target are shown
as the solid circles and counts from the carbon target
(scaled by the ratio of integrated luminosities on 
bound nucleons) are shown as the open circles. 
All other relevant exclusivity cuts have been applied.
}
\label{fig:wwwep}
\end{figure}

In the case where there is a proton measured in the
final state, the electron-proton missing mass \mxep
should equal the pion mass, with the assumption of quasi-free
production. Distributions 
for the only relevant topology, \tpIV are shown
in Fig.~\ref{fig:wwwep}a for Part 1p6i
and Fig.~\ref{fig:wwwep}b for Part 4p2o. 
The background carbon distributions are rather similar
to those from the ND$_3$ target, aside from a slight
shift due to the higher average binding energy in
$A>2$ nuclei, so that only a modest reduction in
background can be achieved. We used a single
set of cuts for all beam energies: 
$-0.11 < (M_x^{eN})^2 < 0.15$ GeV$^2$.

\subsubsection{Missing Energy Cut}
In the topology \tpIVn,
the energy of all the final state particles is measured,
and therefore the missing energy $E_m$ distribution should be 
centered around 2 MeV, assuming quasi-free production
from a deuteron. If
the event came from a nucleus with $A>2$, such as helium, 
nitrogen, or aluminum, the missing energy will be 
larger, of order 20 MeV, due to the typical binding
energy of a nucleon in a nucleus. 
Unfortunately the energy resolution of CLAS
is not sufficient to clearly distinguish between
quasi-free and bound nucleons, but nevertheless we did find
that placing an upper cut made a small
improvement in the signal-to-background ratio. 
The cut $E_{m}<0.065$ GeV
was used for all kinematic bins and all Parts.

\subsubsection{Angular Cuts}
\begin{figure}[hbt]
\centerline{\includegraphics[height=7.0in,angle=90]{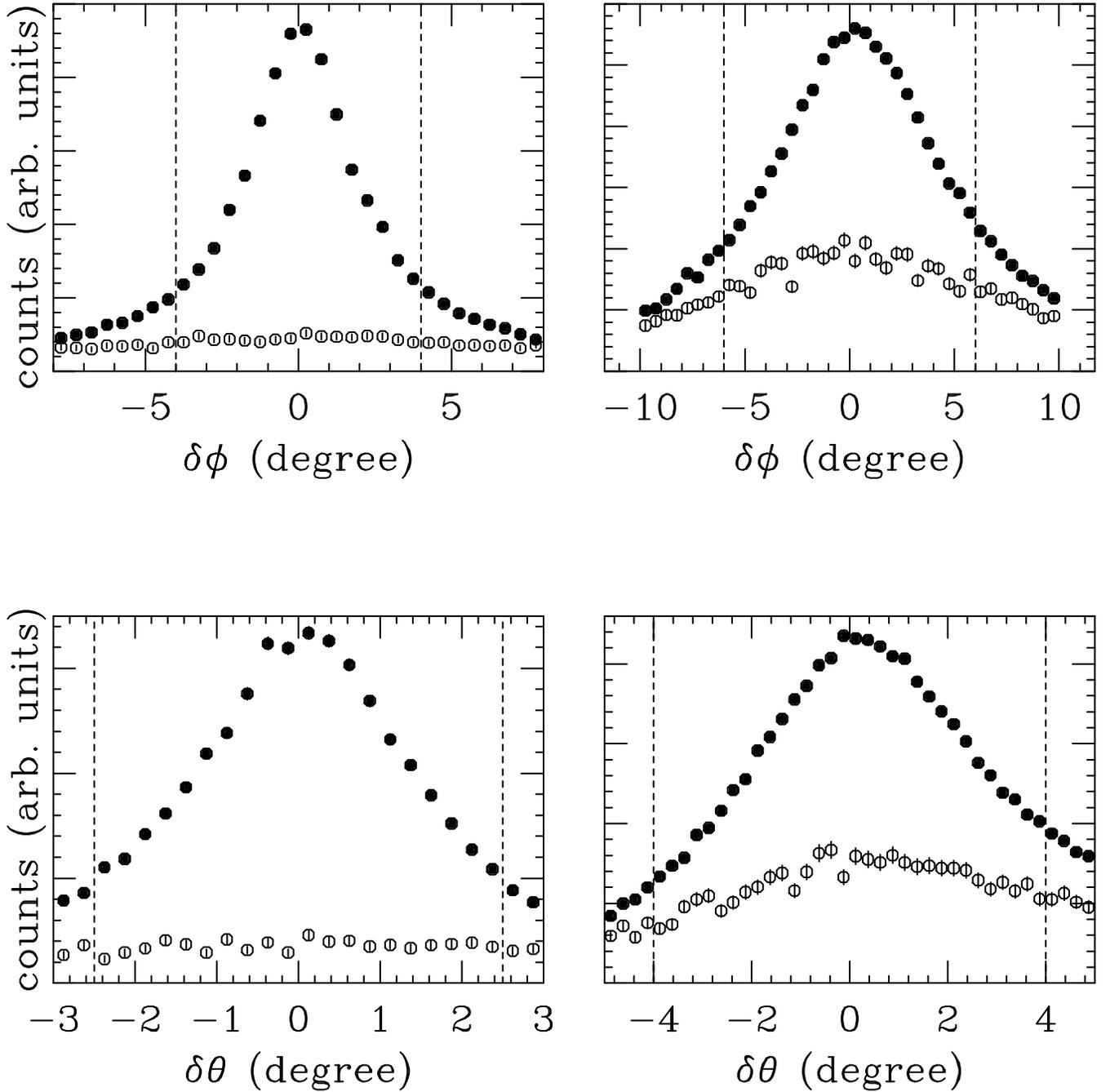}}
\caption{Distributions of angular differences in the
predicted and observed nucleon direction cosines from Part 4p2o for
topology \tpI (left panels) and topology \tpIV (right panels). 
The top row is for $\delta\phi$
and the bottom row is for $\delta\theta$. 
The black points are for the ammonia target, while
the open circles are from the carbon target, scaled
by integrated luminosity. The vertical dashed lines
indicate the cuts used in the analysis. 
All other relevant exclusivity cuts have been applied.
}
\label{fig:thphi}
\end{figure}

In the two topologies where all final state particles
are detected, angular cuts are very useful in 
rejecting background from $A>2$ materials. From the
kinematics of the detected electron and meson, the
direction cosines of the recoil nucleon are calculated,
and compared with the observed angles. We denote
the difference in predicted and observed angles
as $\delta \theta$ in the in-plane direction and
$\delta \phi$ in the out-of-plane direction (which
tends to have worse experimental resolution). Distributions
of these two quantities are shown for the relevant
topologies in Fig.~\ref{fig:thphi}, averaged over
all kinematic variables, for Part 4p2o. 
The dashed lines show
the cuts used to optimize the signal-to-background ratio. The
kinematic dependence of the angular resolution was found
to be sufficiently weak to justify the use of a single cut
value for all kinematic values.
The cut values are listed in Table~\ref{tab:thphi}.

\begin{table}[hbt]
\begin{tabular}{lll}
topology & \> $\delta\theta$ cut& \> $\delta\phi$ cut \\
\hline 
\tpI    & \> $|\delta\theta|<2.5^\circ$ & \> $|\delta\phi|<4^\circ$ \\
\tpIV   & \> $|\delta\theta|<4^\circ$ & \> $|\delta\phi|<6^\circ$ \\
\end{tabular}
\caption{Cuts on $\delta\theta$ and $\delta\phi$
for each of the relevant topologies.
}
\label{tab:thphi}
\end{table}

\subsection{Kinematic Binning}\label{sec:kin}

\begin{figure}[hbt]
\centerline{\includegraphics[height=7.0in,angle=90]{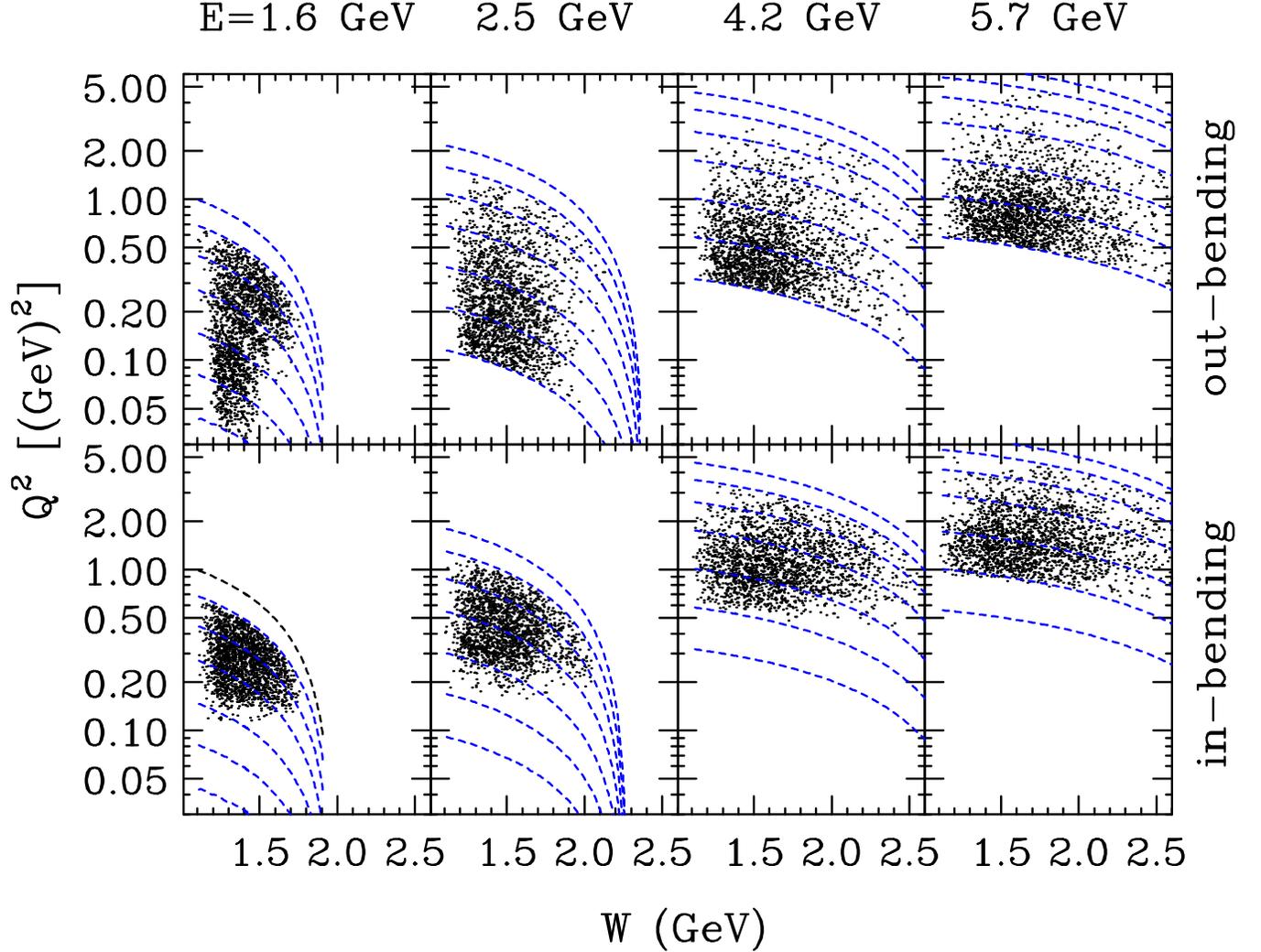}}
\caption{(color online) Distribution in $(W,Q^2)$ of events for the \tpI
topology passing all exclusivity cuts, for four different
beam energies, from left to right. The top (bottom) row of panels are
for the out-bending (in-bending) torus polarity for
negatively charged particles.  The
blue dashed lines show 
the bin limits in $Q^2$, defined by fixed bins in $\theta_e$.
}
\label{fig:wq2}
\end{figure}

The kinematic
range of the experiment is $1.1<W<2.6$ GeV and 
$0.05<Q^2<5$ GeV$^2$. As shown in Fig.~\ref{fig:wq2},
the range in $Q^2$ changes with $W$. We therefore
made six bins in $Q^2$, where the
limits correspond to electron scattering angles
of 7.5, 10.0, 13.3, 17.6, 23.4, 31.0, and 42.0 degrees. In order to
study possible resonance structure in the \tpI reaction,
we used nominal $W$ bins of width 0.03, 0.04, or 0.05 GeV
for beam energies near 1.7, 2.5, and 4.2 to  5.7 GeV,
respectively. The bin widths increase
slightly for $W>2$ GeV. These bin sizes are comparable
to the experimental resolution. For the \tpIV reaction,
we used $W$ bin widths that are three times larger 
than for the \tpI reaction (i.e. 0.09, 0.12, or 0.15 GeV).
This sacrifice was made in order that the majority
of bins had at least 10 counts (the minimum needed 
for Gaussian statistical uncertainties).

An examination of event rates showed a strong forward
peaking in \cthcmsp for all the topologies
studied, roughly independent of $(W,Q^2)$. 
We use twelve bins in $\cos(\theta^*)$, with
boundaries at -1.0, -0.8, -0.6, -0.4, -0.2, 0, 0.2, 
0.4, 0.6,  0.7, 0.8, 0.9, and 0.995. Finer bins in \cthcmsp
were used for $\cos(\theta^*)>0$ because the cross sections
tend to be forward-peaked, especially at the higher values
of $W$ and $Q^2$. Because the pion polar angle was limited to
48 degrees in the lab frame, most of the bins with 
$\cos(\theta^*)<0$ are empty.
The upper-most boundary of 0.995 was chosen instead of 1.0
because the average
resolution in \phicmsp becomes worse than 30 degrees above
$\cthcmn=0.995$, making it increasingly problematic to determine the
$\phi^*$-dependence of spin asymmetries at very forward angles.
 
We use 12 bins in $\phi^*$, equally spaced
between 0 and $2\pi$. We chose 12 bins in order to
be able to distinguish between terms proportional
to $\sin(\phi^*)$ and those proportional to $\sin(2\phi^*)$.

For most bins, the average
values of $(W,Q^2,\cos(\theta^*),\phi^*,\epsilon)$ are
very close to the bin centers. No bin-centering corrections
were applied to the data. Instead, the count-weighted average values
of all relevant kinematic variables are included in the
table of results~\cite{clasdb,SMp,SMm}.

The values of $(W,Q^2,\cos(\theta^*),\phi^*,\epsilon)$ were obtained
assuming the that struck nucleon is at rest, which is a valid assumption
for the \tpI reaction, but not the \tpIV reaction. In the latter case,
the typical momentum of the struck nucleon is of order 0.05 GeV, 
which introduces an uncertainty in $(W,Q^2,\cos(\theta^*),\phi^*)$
that is less than, or in the worst case, comparable to the 
chosen bin sizes for this reaction. 

\subsection{Asymmetries}
Spin asymmetries were formed as follows:
\be
A_{LL} = \frac{
   N^{\uparrow\downarrow} + 
   N^{\downarrow\uparrow} -
   N^{\uparrow\uparrow} - 
   N^{\downarrow\downarrow}}{
   N_{tot} \hskip .05in f \hskip .05in P_BP_T},
\ee
\be
A_{UL} = \frac{
   N^{\uparrow\uparrow} + 
   N^{\downarrow\uparrow} -
   r_TN^{\uparrow\downarrow} - 
   r_TN^{\downarrow\downarrow}}{
   N_{tot} \hskip .05in f \hskip .05in P_T},
\ee
where the symbols $N$ represent the number of events
in a given helicity configuration, divided by the
corresponding integrated beam current. The first superscript
refers to the beam polarization direction and the
second to the target polarization direction. Here 
 $N_{tot}=N^{\uparrow\uparrow} + 
               N^{\downarrow\uparrow} +
               N^{\uparrow\downarrow}r_T + 
               N^{\downarrow\downarrow}r_T$,
and $f$ is the dilution factor, defined as the fraction
of events originating from polarized nucleons compared
to the total. The factor $r_T$ is the ratio of
effective target thicknesses for data taken with positive and
negative target polarization (reversal frequency typically
three days) and ranged from 0.98
to 1.02, except for the deuteron runs of Part 5p7i, where
the correction was 0.94. No correction was needed
for running with positive and negative beam helicity
due to the rapid (30 Hz) reversal rate.
The target polarization $P_T$ is the luminosity-weighted
average of the absolute value of the positive and
negative target polarization data. The effect of the small
difference in absolute value of negative and positive
target polarizations is taken into account through our
method of determining $P_BP_T$ from $ep$ elastic
(quasi-elastic) scattering.  

The sign convention for $A_{LL}$ corresponds to 
a positive value when the cross section for
scattering to a spin $S=\frac{1}{2}$ final state is larger than to a 
$S=\frac{3}{2}$  final state (see Sec.~\ref{sec:formalism}).

\subsection{Beam and Target Polarization}
\label{sec:pbpt}
The product of beam 
polarization ($P_B$) and target polarization ($P_T$) was
determined using the well-understood beam-target spin
asymmetry in elastic $ep$ scattering (quasi-elastic scattering
for the deuteron target). The results~\cite{inclp,incld}
are listed in Table~\ref{tab:Parts}. 
The beam polarization
was measured using M\o ller scattering. The average value
was 0.70, with a spread of about 4\% (relative). No 
dependence on incident beam energy was observed.
For determining the target spin asymmetry $A_{UL}$,  
the proton or deuteron target polarization was 
determined by dividing the values of $P_BP_T$ from
$ep$ elastic scattering by $P_B$ from the M\o ller
measurements. 

\subsection{Combining Similar Parts}
The number of events as well as the average value of kinematic
quantities was stored for each kinematic bin for each Part separately.
Before extracting the dilution
factor and asymmetries, the counts and averaged kinematic quantities
were combined for  
Parts 5p6i and 5p72i and the result is named Part 5p7i.

\subsection{Dilution Factor}
\label{sec:df}
The dilution factor $f$ is defined as the ratio of 
scattering rate
from free nucleons to the scattering rate from all nucleons in the target.
If we make the assumption that the cross section per nucleon
is the same for bound protons in all of the nuclear
materials (with $A>2$)~\cite{Adep}
 in a given target, and also that
the effective detection efficiency is the same for the
ammonia and carbon targets, then
\be
f = 1 - R_{A>2} \frac{N_C}{N_{NX_3}},
\label{Eq:f}
\ee
where 
$N_C${ and $N_{NX_3}$ are the number of counts
measured in a given kinematic bin for a given topology,
normalized by the corresponding integrated beam charge,
and $R_{A>2}$ is the ratio of the number of bound nucleons
in the  ammonia target to the number of bound nucleons in the
carbon target. Bound nucleons are defined to be in
materials with atomic number $A>2$.
The latter was determined from a detailed analysis of the
target composition using inclusive electron scattering
rates from ammonia, carbon, and 
empty targets~\cite{inclp,incld}.
The ratio must be determined separately for bound
protons in the NH$_3$ target (for the \tpI reaction)
and for bound neutrons in the ND$_3$ target
(for the \tpIV reaction). We denote these ratios as
$R_{A>2}^p$ and $R_{A>2}^d$, respectively, and list the
values used in the analysis in Table~\ref{tab:Parts}.
Using a study of inclusive electron scattering rates, we
found $R_{A>2}^p$ to vary between 0.81 and 0.86 
for the various Parts of the
experiment, while $R_{A>2}^d$ varied between 0.99 and 1.01.
The variation
is due to the target material being replaced periodically
during the experiment, and also due to settling of the 
ammonia beads.

Because the integrated luminosities on the carbon target
were generally about five times lower than on the ammonia
targets, there is a large amplification of the uncertainty
on the ratio of carbon to ammonia counts, 
$\frac{N_C}{N_{NX_3}}$. In many cases, this would lead
to unphysical values of $f$ (i.e. $f<0$). We therefore
took advantage of the fact that $f$ is a very slowly
varying function of kinematic variables, and did a global fit
to $\frac{N_C}{N_{NX_3}}$ for each topology and run configuration. 
The fit values were then
used to evaluate $f$ in each kinematic bin.

 Several functional forms for the fits to 
$\frac{N_C}{N_{NX_3}}$ were tried. The final form
selected was:
$$
\frac{N_C}{N_{NX_3}}  = P_1[1 \, + \, 
P_2 W \, + \, P_3 Q^2 \, + \, P_4 \cthcmn \, + \,
P_5 W^2 \, + \, P_6 W Q^2 \, +$$
$$ \, P_7 W \cthcmn \, + \, \\
P_8 (Q^2)^2 \, + \, P_9 W Q^2\cthcmn \, + 
\, P_{10} \cthcmn^2 \, + \, 
 P_{11} R_1(W^2) \, + \,
P_{12} R_2(W^2) \, + \,
P_{13} R_3(W^2) \, +$$
$$P_{14} W^2 \cthcmn \, +
P_{15} R_1(W^2)  \cthcmn +
P_{16} R_2(W^2)  \cthcmn +
P_{17} R_3(W^2)  \cthcmn + $$
$$
P_{18} R_2(W^2)  \cthcmn Q^2 +
P_{19} R_3(W^2)  \cthcmn Q^2 +
P_{20} R_2(W^2)  \cthcmn^2 + $$
$$
P_{21} R_3(W^2)  \cthcmn^2 +
P_{22} R_2(W^2)  Q^2 +
P_{23} R_3(W^2)  Q^2 + $$
$$ R_4(W^2) (P_{24} + P_{25}\cthcmn + P_{26} Q^2 +    
             P_{27} \cthcmn Q^2 + P_{28}\cthcmn^2) + $$
$$ R_5(W^2) (P_{29} + P_{30}\cthcmn + P_{31} Q^2 +    
             P_{32} \cthcmn Q^2 + P_{33}\cthcmn^2) 
],$$

\noindent where the functions 
\be
R_i(W^2)=\frac{\Gamma_i}{(W^2 - W_i^2)^2 + (W_i \Gamma_i)^2}
\ee
are to account for the influence
of the five prominent resonances~\cite{maid} at 
$W_1=1.23$ GeV, 
$W_2=1.53$ GeV,     
$W_3=1.69$ GeV,
$W_4=1.50$ GeV,
$W_5=1.43$ GeV, and
with widths 
$\Gamma_1=0.135$ GeV, 
$\Gamma_2=0.220$ GeV, 
$\Gamma_3=0.120$ GeV, 
$\Gamma_4=0.080$ GeV, and 
$\Gamma_5=0.370$ GeV. The reason that these resonance terms are
needed is that the nucleon resonances are very much broadened
in the target materials with $A>2$, but have the natural width
for free nucleons. This generates resonant-like structures in
the ratio of carbon to ammonia count rates. 
The other terms are simply
power-law expansions in terms of $W$, $Q^2$, and \cthcm. 

All of the fit parameters were used for the 
highest-statistical accuracy topologies \tpV and \tpVIIIn. 
For the low count rate topologies \tpI and \tpIVn,
parameters 8-10 and 14-33 were fixed at
zero. Tests were made to see if any $\phi^*$-dependent
terms would improve the fits. No significant 
improvements were found. 
Comparisons of fits and data are 
made for the two $\pi^+$ topologies in 
Fig.~\ref{fig:dil65} for Part 4p2o. The data show considerable
resonance structure, and this Part was the most difficult
in order to obtain a good fit. 

%todo fix up figure
\begin{figure}[hbt]
\centerline{\includegraphics[height=7.0in,angle=90]{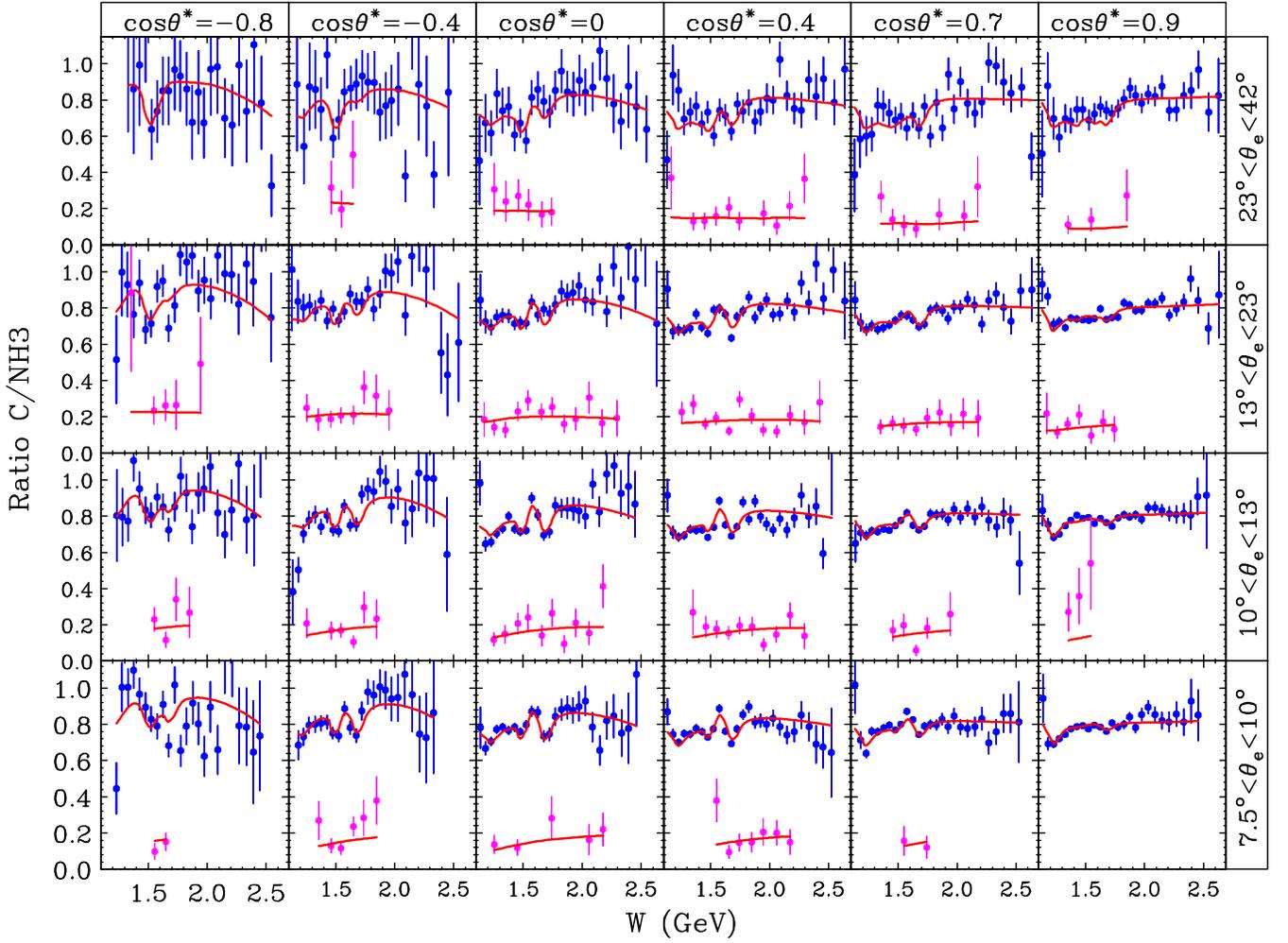}}
\caption{(color online) Ratios of count rates from the carbon target  to
count rates from the NH$_3$ target for Part 4p2o
of the experiment, for events passing all relevant
exclusivity cuts.
The panels correspond from left to right to
six \cthcmsp bins and from bottom to
top to four ranges of $\theta_e$.
The larger sets of ratios, shown in blue, correspond
to the topology \tpVn, while the smaller values,
shown in magenta, 
correspond to the fully exclusive topology \tpIn. 
The red curves are the fits to the data described in the
text (with the upper curves matching the blue points,
and the lower curves matching the magenta points). 
}
\label{fig:dil65}
\end{figure}

The dilution factors were evaluated using Eq.~\ref{Eq:f} and the fits to
$\frac{N_C}{N_{NX_3}}$. The results for
the two $\pi^+$ topologies are shown for Part 4p2o in 
Fig.~\ref{fig:dilf65} as a function of $W$ in a grid
over $\theta_e$ and $\cthcmn$. 
For the fully
exclusive topology, \tpIn, the dilution factor
is very high, about 0.9 on average, corresponding
to the good rejection of background that is possible
with the exclusivity cuts when the recoil neutron
is detected.  For the topology \tpVn, the dilution 
factor is reasonably good, averaging about 0.4, 
with some oscillations due to resonance structure.
At central values of \cthcm, the resolution in 
electron-pion missing mass is poor, especially at
high beam energies, causing the dilution factor to
drop to 0.2 for some bins. 

\begin{figure}[hbt]
\centerline{\includegraphics[height=7.0in,angle=90]{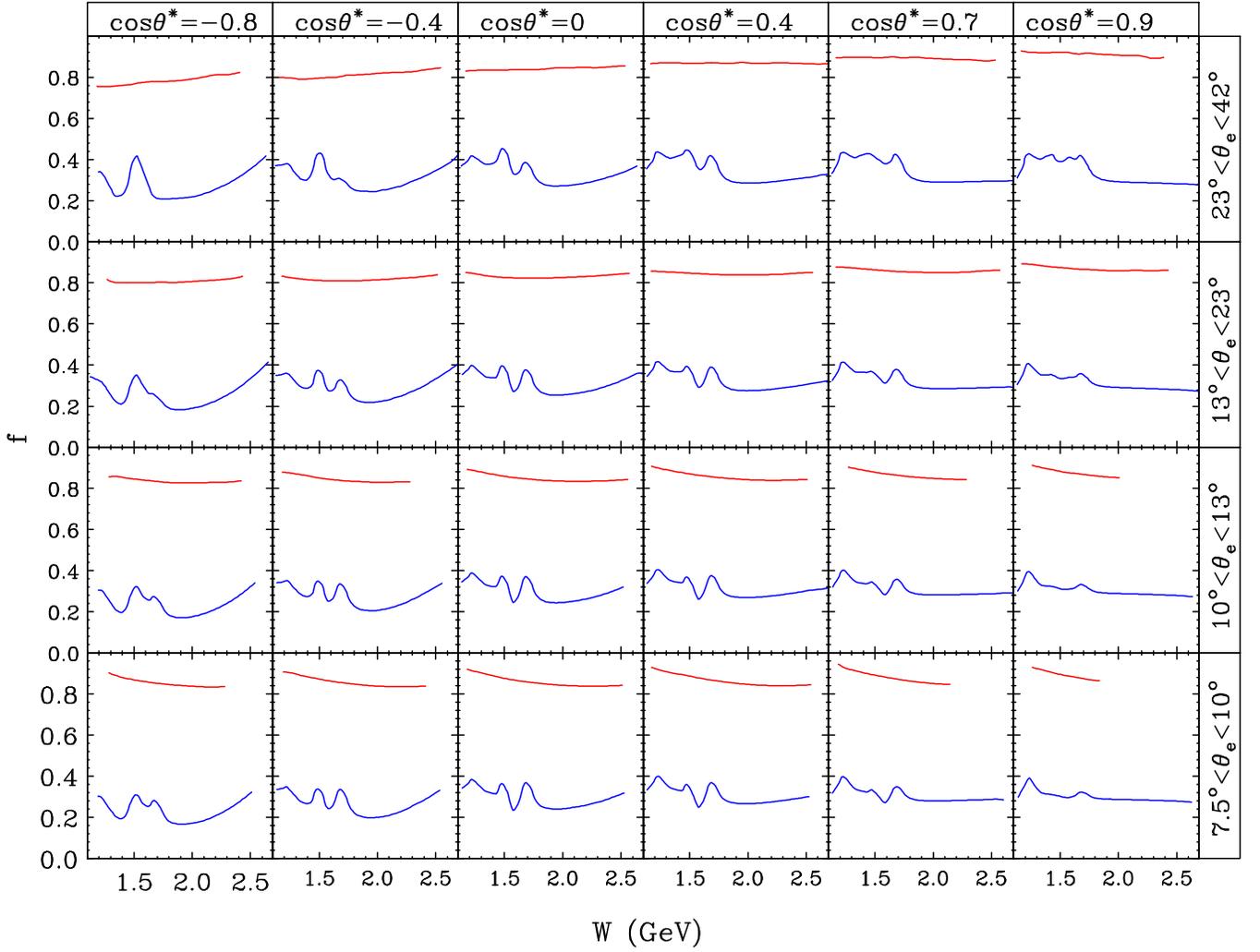}}
\caption{(color online) Dilution factors as a function of $W$ for
the two $\pi^+$  topologies for Part 4p2o in six \cthcmsp bins
(from left to right) and four $\theta_e$
bins (from bottom to top). The upper red curves within each
panel correspond
to topology \tpI and the lower blue curves to topology
\tpVn. 
}
\label{fig:dilf65}
\end{figure}

The dilution factors for the two $\pi^-$ topologies 
are shown in Fig.~\ref{fig:dilf68} for Part 4p2o.
For the fully exclusive topology, the dilution factor is
quite good, averaging around 0.8. The topology with
a missing proton, \tpVIIIn, has a much worse dilution
factor, which is more than compensated for by a much
higher event rate. The exception is at backward angles
in the center-of-mass, where the dilution factor
falls below 0.1 for the higher beam energies. In the worst
cases, it is actually close to zero, implying no exclusive
signal at all, compared to the very large backgrounds
as \cthcmsp approaches -1.
Asymmetry results were not evaluated for any kinematic bins
for which the dilution factor was less than 0.1,
due to the increasingly divergent uncertainty on the dilution factor.

\begin{figure}[hbt]
\centerline{\includegraphics[height=7.0in,angle=90]{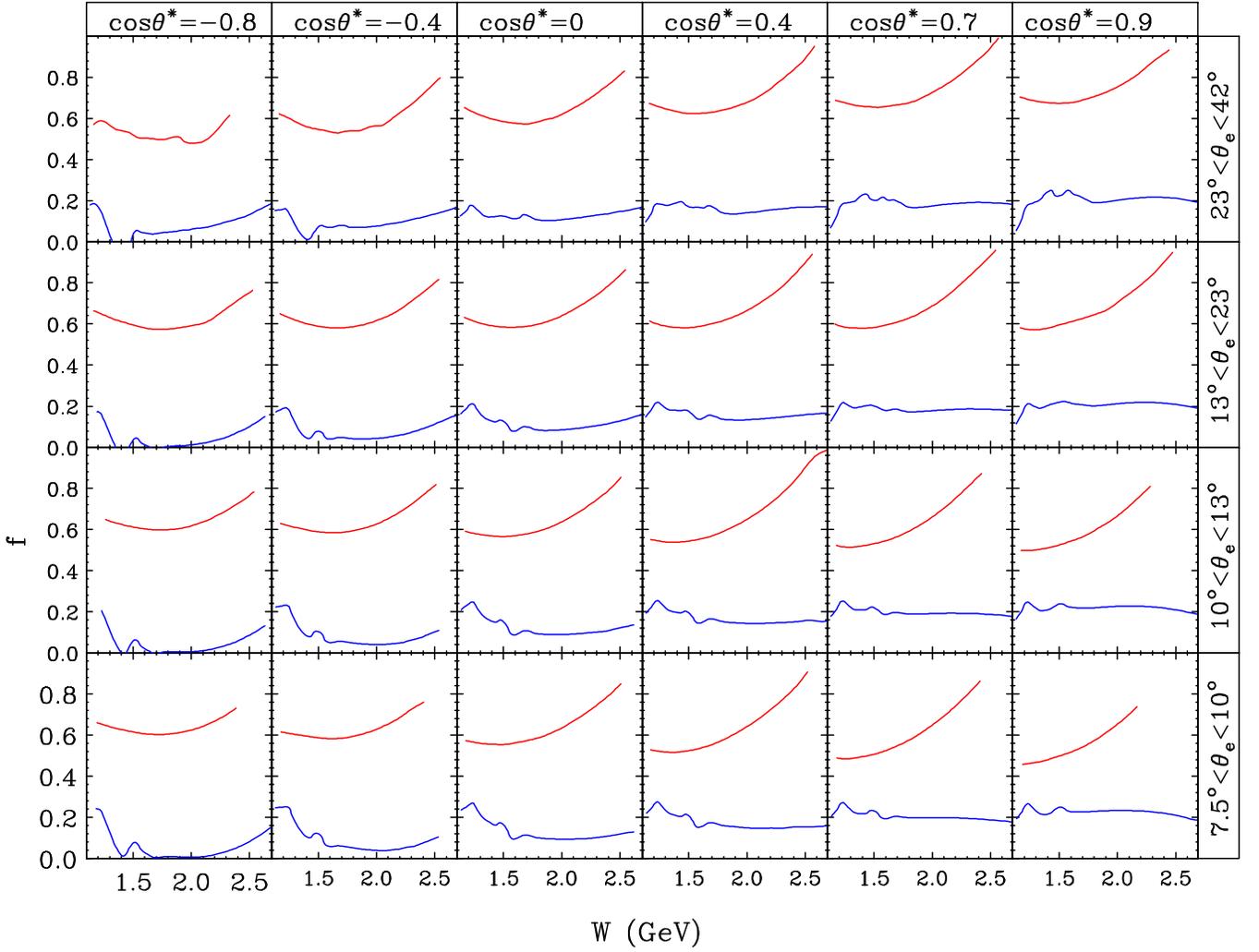}}
\caption{(color online) Same as Fig.~\ref{fig:dilf65}, except for
the two $\pi^-$ topologies: \tpIV (upper red curves)
and \tpVIII (lower blue curves). 
}
\label{fig:dilf68}
\end{figure}

\subsection{Radiative Corrections}
An extensive
study of radiative corrections to exclusive longitudinal
spin asymmetries was performed using the equivalent radiator
and angle-peaking approximations of the well-known
Mo-Tsai formalism~\cite{tsai}. Although radiative corrections
are very important for the extraction of 
cross sections, they were found to be negligible for 
spin asymmetries (less than 0.5\%).

\subsection{Polarized Nitrogen Correction}
The $^{15}$N isotope
in the ammonia targets is slightly polarized, with a scale
factor of about -0.018 
relative to the free protons~\cite{E143}.
In the present exclusive analysis,
the correction to the reaction \tpI is reduced to a smaller level
(on average about -0.003 for topology \tpI and -0.009 for topology
\tpVn) because most of the events from nitrogen are removed
by the exclusivity cuts. Because of the theoretical 
uncertainty in evaluating the corrections, they were not
applied to the data, but rather treated as a systematic
uncertainty. In the
absence of D-state and final state interaction corrections, the
correction to the reaction \tpIV is negligible, relative to
other systematic uncertainties.

\subsection{Combining Data Sets}
The asymmetry analysis was performed for each topology and
each Part separately (see Table~\ref{tab:Parts}).
Since the in-bending and out-bending
Parts had similar or identical beam energies, the asymmetries
should be the same in a given kinematic bin.
Specifically, we combined the following parts: (1pti, 1p7o), 
(2p2i, 2p6i, 2p5o), (42i, 4p2o), and (5p6i, 5p72i, 5p72o). 
The inbending Parts favored larger scattering angles, while
the outbending Parts went to much smaller electron scattering angles.
Combining the two together gives a relatively uniform
coverage in $\theta_e$ (and hence $Q^2$).

The configurations were combined by adding 
asymmetries together in quadrature for each of the
4-dimensional bins. Since the two configurations
differ only in the acceptance function, which should
cancel in forming asymmetries, the expectation is
that they should be fully compatible statistically.
This was verified by forming the $\chi^2$ per 
degree of freedom for combining each of the two
asymmetries, for each of the four topologies, and for
the four beam energies. As
can be seen in Table~\ref{tbl:chisqc}, the in-bending
and out-bending 
configurations indeed are consistent with each other.
The weighted average of all kinematic quantities was taken
when combining the configurations. 

\begin{table}[hbt]
\begin{tabular}{llcccc}
$E_0$         & & $A_{LL}$ & & $A_{UL}$ &  \\
(GeV) & topology & $\chi^2/$d.f. & d.f. &
$\chi^2/$d.f. & d.f.  \\
\hline 
  1.7 & \tpIV   &    0.91 &    219 &    1.18 &    219\\
  1.7 & \tpV    &    1.01 &   5001 &    1.13 &   4294\\
  1.7 & \tpVIII &    1.03 &   1679 &    1.03 &   1679\\
  2.5 & \tpI    &    1.00 &    162 &    1.14 &    160\\
  2.5 & \tpIV   &    1.00 &    588 &    1.08 &    588\\
  2.5 & \tpV    &    1.04 &   7204 &    1.10 &   7197\\
  2.5 & \tpVIII &    1.00 &   2893 &    1.07 &   2893\\
  4.2 & \tpI    &    1.00 &    310 &    1.12 &    310\\
  4.2 & \tpIV   &    1.07 &    585 &    1.08 &    582\\
  4.2 & \tpV    &    0.97 &   6799 &    1.10 &   6796\\
  4.2 & \tpVIII &    0.98 &   2113 &    1.05 &   2107\\
  5.7 & \tpI    &    1.05 &    110 &    1.15 &    110\\
  5.7 & \tpIV   &    0.97 &    207 &    0.99 &    207\\
  5.7 & \tpV    &    0.99 &   4993 &    1.09 &   4993\\
  5.7 & \tpVIII &    1.03 &   1314 &    1.09 &   1313\\
\end{tabular}
\caption{Values of $\chi^2$/d.f. and number of 
degrees of freedom (d.f.) for combining both
asymmetries from the inbending and outbending torus polarity
Parts of each beam energy range.
}
\label{tbl:chisqc}
\end{table}

\subsection{Combining Topologies}

For both positive and negative pion electroproduction,
we combined the fully exclusive topology with the one
with one missing hadron. This was done by forming
a weighted average of the two results on a bin-by-bin
basis. 
For both the $\pi^+$ and $\pi^-$ final states, and for
both asymmetries, the
topologies were found to be statistically compatible, as shown
by the good values of $\chi^2$/d.f. in
Table~\ref{tbl:chict}. 
In forming $\chi^2$, 
each degree of freedom corresponds to an individual
point in $(W,Q^2,\cos(\theta^*),\phi^*)$ for which
both topologies had at least 10 raw counts.

\begin{table}[hbt]
\begin{tabular}{llcccc}
$E_0$ &         & $A_{LL}$ & & $A_{UL}$ &  \\
(GeV) & reaction & $\chi^2/$d.f. & d.f. &
$\chi^2/$d.f. & d.f.  \\
  1.7 & \tpI    &    1.01 &    433 &    1.13 &    428\\
  1.7 & \tpIV   &    0.94 &    707 &    1.26 &    707\\
  2.5 & \tpI    &    1.00 &   1022 &    1.16 &   1000\\
  2.5 & \tpIV   &    1.00 &   1555 &    1.20 &   1553\\
  4.2 & \tpI    &    1.05 &   1339 &    1.13 &   1336\\
  4.2 & \tpIV   &    1.04 &   1588 &    1.06 &   1576\\
  5.7 & \tpI    &    1.02 &    663 &    1.18 &    662\\
  5.7 & \tpIV   &    0.89 &    628 &    1.11 &    628\\
\hline 
\end{tabular}
\caption{Values of $\chi^2$/d.f. and number of 
degrees of freedom (d.f.) for combining the two
asymmetries from the topologies with all particles
detected with the topology with a missing nucleon.
}
\label{tbl:chict}
\end{table}

\subsection{Systematic Uncertainties}
The systematic uncertainty in the asymmetry results
is dominated by overall scale factor uncertainties
arising from the uncertainties in the beam and target
polarizations, and from the uncertainty in the dilution
factor, as shown in Table~\ref{tbl:syst}. More details
on each of the contributing factors are given in the
next sub-sections. 

\subsubsection{Target and Beam Polarization}
The product of beam and target polarization was determined
for the polarized proton target from $ep$ elastic events 
with a relative statistical precision ranging from 
1\% at low beam energies to about 3\% at 5.7 
GeV~\cite{inclp,incld}. A
spread of about 1.5\% was observed in comparing the
results with different event selection criteria. These 
quantities were combined in quadrature for the net uncertainty
on $P_BP_T$. The relative uncertainties for the deuteron are much
larger than for the proton, 
principally because the average target polarization
is almost three times smaller for the deuteron than for the proton.

The uncertainty on the beam polarization was estimated to 
be 4\%~\cite{inclp}. 
We combined the uncertainty on $P_BP_T$ and the uncertainty on $P_B$
in quadrature to determine the uncertainty on $P_T$ itself.

\begin{table}[hbt]
\begin{tabular}{l | c c | c c | c c | c c}
 \multicolumn{9}{c}{ \tpI} \\
   \multicolumn{1}{c}{       } &
   \multicolumn{2}{c}{1.6 GeV} &
   \multicolumn{2}{c}{2.5 GeV} &
   \multicolumn{2}{c}{4.2 GeV} &
   \multicolumn{2}{c}{5.7 GeV} \\
\hline
quantity & $A_{LL}$ &            $A_{UL}$ &
           $A_{LL}$ &            $A_{UL}$ &
           $A_{LL}$ &            $A_{UL}$ &
           $A_{LL}$ &            $A_{UL}$ \\
$P_B,P_T$ & 0.02 &        0.04 & 
            0.02 &        0.05 &
            0.03 &        0.05 &
            0.03 &        0.05 \\
$f$       & 0.03 &        0.03 &
            0.04 &        0.04 &
            0.05 &        0.05 &
            0.06 &        0.06 \\
$^{15}N$  & 0.01 &        0.01 &
            0.01 &        0.01 &
            0.01 &        0.01 &
            0.01 &        0.01 \\
Total     & 0.04 &        0.05 &
            0.04 &        0.06 &
            0.05 &        0.06 &
            0.06 &        0.08 \\

\hline
 \multicolumn{9}{c}{ } \\
 \multicolumn{9}{c}{ \tpIV} \\
   \multicolumn{1}{c}{       } &
   \multicolumn{2}{c}{1.6 GeV} &
   \multicolumn{2}{c}{2.5 GeV} &
   \multicolumn{2}{c}{4.2 GeV} &
   \multicolumn{2}{c}{5.7 GeV} \\
\hline
quantity & $A_{LL}$ &$       A_{UL}$ &
           $A_{LL}$ &       $A_{UL}$ &
           $A_{LL}$ &       $A_{UL}$ &
           $A_{LL}$ &       $A_{UL}$ \\
$P_B,P_T$ & 0.05 & 0.06 & 0.07 & 0.07 & 0.10 & 0.10 & 0.15 & 0.15 \\
$f$       & 0.10 & 0.10 & 0.12 & 0.12 & 0.15 & 0.15 & 0.20 & 0.20 \\
$^{15}N$  & 0.02 & 0.02 & 0.02 & 0.02 & 0.02 & 0.02 & 0.02 & 0.02 \\
Total     & 0.11 & 0.12 & 0.14 & 0.14 & 0.18 & 0.18 & 0.25 & 0.25 \\
\hline 
\end{tabular}
\caption{Estimated relative scale uncertainties for
the various beam energies and asymmetries of the
experiment from beam/target polarization, from
dilution factor $f$, and from polarized nitrogen.
}
\label{tbl:syst}
\end{table}

\subsubsection{Dilution Factor}
The systematic uncertainty on the dilution factor arises from
four factors. The first is how well the multi-parameter
fit describes the measured ratios of rates from the carbon and
ammonia targets. From the reasonably good values of
$\chi^2/d.f.$ for the fits, we conclude that all of the
 significant resonance structures in the ratios are
accounted for by the fits at the few percent level.
It is also possible for there to be a $\phi^*$ dependence to
the ratios, although the fits were not improved when we
included terms proportional to $\cos(\phi^*)$. 

The second source of uncertainty is in the factors  $R_{A>2}^p$
and $R_{A>2}^d$, defined as the ratio of protons
(neutrons) in target materials with $A>2$ for the 
ammonia target compared to the carbon target. We compared
three methods of determining these factors: a study of
inclusive electron scattering rates; fits to the
electron-pion missing mass spectra for values well below the
nucleon mass; and the value that gives
the best agreement for $A_{LL}$ between the fully
exclusive topologies and the topologies where the
recoil nucleon is not detected. This last technique
relies on the fact that the fully exclusive topologies
have much less nuclear background. From these comparisons,
we estimate a typical systematic uncertainty of about 2\% (relative)
for $R_{A>2}^p$ and 4\% for $R_{A>2}^d$. From Eq.~\ref{Eq:f}, 
this translates, on average, 
into approximately 4\% (12\%) overall normalization uncertainty on
the \tpI (\tpIV) asymmetries $A_{LL}$ and $A_{UL}$. 
We found the systematic uncertainty to increase with increasing
beam energy, due to the limited accuracy with which
the three methods could be compared at higher beam energies.

The third potential source of uncertainty comes from the fact
that the carbon target contained about 20\% more helium than the 
ammonia targets. If the ratio of helium to carbon has a significant
kinematic dependence, it could translate into a variation of
the dilution factor with kinematic variables, relative to the
average correction. We examined the ratio of ``empty target'' 
(mostly helium)  to carbon
target rates within the standard cuts of the highest statistical
accuracy topology, \tpV, and found variations of less than 
$\pm5\%$, corresponding to an uncertainty of about $\pm 1$\% in $f$.

A fourth source of uncertainty could arise from a difference
in the Fermi broadening in $^{15}$N compared to $^{12}$C,
or a difference in average binding energy. In order to
place constraints on this possibility, a dedicated liquid
$^{15}$N target was built for the present experiment,
and inclusive electron scattering rates were compared
with those from carbon~\cite{Adep}. Within the
limited statistical and systematic accuracy of the
measurements (the latter being dominated by the
uncertainty in the neutron-to-proton cross section ratio),
the average Fermi momentum and binding energy of the
two nuclei were found to be the same.

\subsubsection{Polarized Nitrogen   }
The systematic uncertainty from the lack of a polarized
nitrogen correction is estimated to be 1\% for the \tpI
reaction~\cite{inclp} (assuming the one-missing-particle
topology \tpV dominates) and at most 1\% for the \tpIV 
reaction~\cite{incld}.

\subsubsection{Multi-Pion Background}
The background from multi-pion production was reduced to
a negligible level due to a combination of two factors.
Firstly, the relatively tight cut on electron-pion
missing mass, which precludes multi-pion background events
unless the electron-proton missing mass resolution
is poor (greater than about 50 MeV). For those few
kinematic bins where the resolution is this poor, the
single-pion peak is so broad that the normalized 
nuclear background is greater than 90\%. These bins were
discarded by the requirement that the dilution factor be
greater than 0.1. Additional constraints come from the
good agreement between the fully exclusive topologies
and the topologies with no recoil nucleon detected. The
former has no multi-pion background due to the many
exclusivity cuts available. 

\subsection{Asymmetries for \tpI from the Deuteron Target}
\label{Sec:pind}
In order to check many aspects of the analysis of the \tpIV
reaction, the \tpI asymmetries from the polarized
proton in the ND$_3$ target were extracted and compared to the 
results from the NH$_3$ target.
This was done for both asymmetries
and all but the highest beam energies. 
%The results for $A_{UL}$ with
%the 2.2/2.5 GeV beam energy are shown in Fig.~\ref{fig:AULpind}.
%The full set of results are shown in Appendix Z. 
The same event selection and exclusivity cuts were used as for
the NH$_3$ target analysis. The same values of beam and target
polarization were used as for the \tpIV analysis. The
dilution factor analysis used the same ratio of nucleons with
$A>2$ in the ND$_3$ target compared to the C target as the
\tpIV analysis, taking into account that it is the number
of protons that is relevant in this case, rather than the
number of neutrons. A comparison of the \tpI reaction from the
two targets (ND$_3$ and ND$_3$) did not reveal any regions of
significant differences beyond those expected from statistical
fluctuations.

% start here but add 
% JANR ref. add cross section refs, add other world data.

\section{Results}
The results of this analysis are tabulated in two large
text files, one for \tpI and one for \tpIVn. Each line in
the table contains the average value of $W$, $Q^2$, \cthcm,
\phicm, $\epsilon$, $\cos(\phi^*)$, $\cos(2\phi^*)$, 
$\sin(\phi^*)$, and $\sin(2\phi^*)$ 
for the particular bin, as well as the
two  asymmetry results along with their statistical uncertainties. 
The systematic uncertainties are negligible in comparison on a
bin-by-bin basis. 
Copies of the tables can be found in the 
CLAS data base~\cite{clasdb} and in the
Supplemental Material associated with this article~\cite{SMp,SMm}.

With approximately 40,000 asymmetry  results for \tpIn, and
15,000 results for \tpIVn, it is a challenge to portray
them in a compact and meaningful way. The variation with
kinematic quantities was examined, and we found very little
dependence on $Q^2$ for a given beam energy, with more
significant variations as a function of $W$ and \cthcm.
There are very strong dependencies on \phicmsp for $A_{UL}$
at all kinematic settings, as well as for $A_{LL}$ at
certain values of $W$ and \cthcm. Based on this study,
the results are presented as a function of \phicmsp 
at each beam energy averaged
over all $Q^2$, adjacent bin pairs in \cthcm, and 
adjacent bin triplets in $W$ for the \tpI reaction.

\subsection{$A_{LL}$ for \tpIn}

The results for $A_{LL}$ for \tpI are shown in 
Fig.~\ref{fig:ALLphipipq1} (1.7 GeV beam energy),
Fig.~\ref{fig:ALLphipipq2} (2.5 GeV),
Fig.~\ref{fig:ALLphipipq3} (4.2 GeV), and 
Fig.~\ref{fig:ALLphipipq4} (5.7 GeV).
Also shown on the plots are the
results of two representative fits to previous data:
the 2007 version of the
MAID unitary isobar fit~\cite{maid} and
the Unitary Isobar version of the JLab Analysis
of Nucleon Resonances (JANR) fit~\cite{janr}, averaged with
the same weighting as the data points. Formally, these two
fits are rather similar in nature, but differ in the data
sets used and in the functional forms used for the 
$Q^2$-dependence of the resonance form factors. 
By and large, both the MAID 2007 and the JANR fits describe the data
reasonably well for
the lowest beam energy. At higher beam energies (and
correspondingly larger values of $Q^2$), both fits
are in reasonably good agreement with data for $W<1.7$ GeV, but 
major differences can be observed at higher values of $W$, with
the magnitude of the differences generally increasing with
increasing $Q^2$.

\begin{figure}[hbt]
\centerline{\includegraphics[height=7.0in,angle=90]{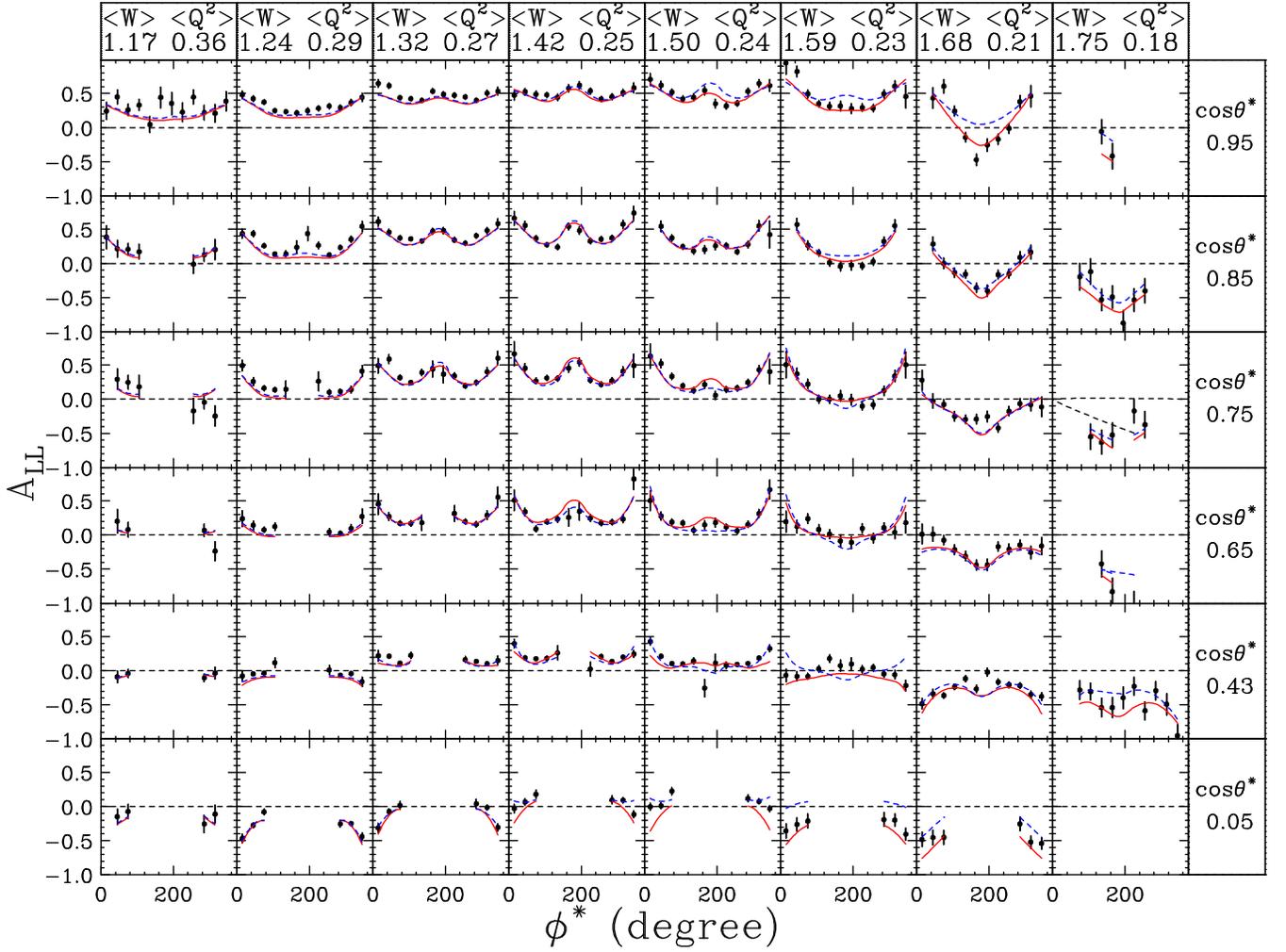}}
\caption{(color online) 
Results for $A_{LL}$ averaged over $Q^2$ as a function of $\phi^*$ 
in eight regions of $W$ (left to right) and the six
regions in \cthcmsp (top to bottom)
for the reaction \tpI and a beam energy range of 1.6 to 1.7 GeV.
The error bars do not include systematic uncertainties. 
The solid red curves are from MAID 2007~\cite{maid} and the
dashed blue curves are from JANR~\cite{janr}. The average values
of $W$ and $Q^2$ are in units of GeV and GeV$^2$ respectively. 
Only results with uncertainties less than 0.2 are plotted, along with
the corresponding model curves. This results in some empty panels. 
}
\label{fig:ALLphipipq1}
\end{figure}

\begin{figure}[hbt]
\centerline{\includegraphics[height=7.0in,angle=90]{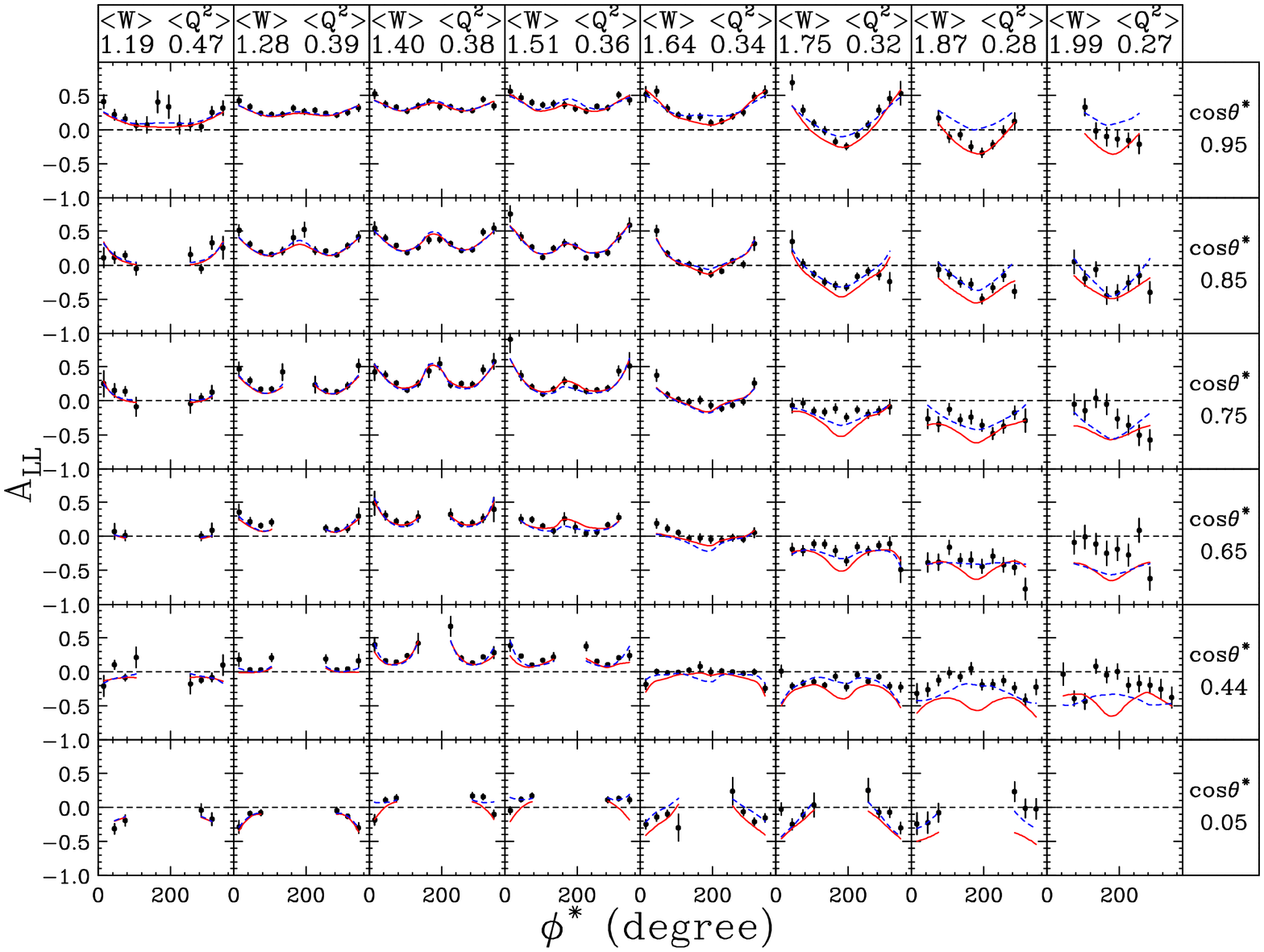}}
\caption{
Same as Fig.~\ref{fig:ALLphipipq1}, except for a beam energy
range of 2.2 to 2.5 GeV. 
}
\label{fig:ALLphipipq2}
\end{figure}

\begin{figure}[hbt]
\centerline{\includegraphics[height=7.0in,angle=90]{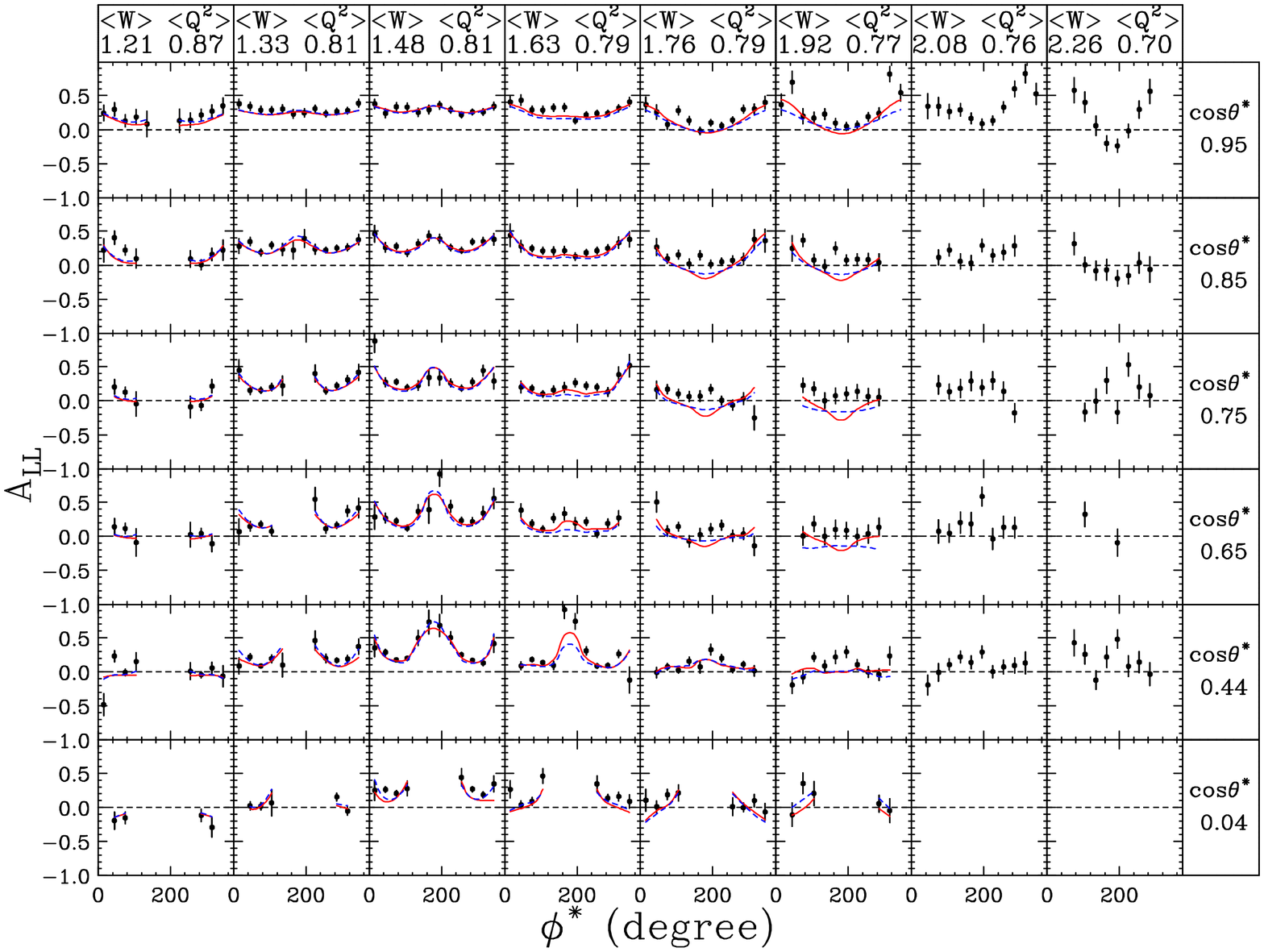}}
\caption{
Same as Fig.~\ref{fig:ALLphipipq1}, except for a beam energy
range of 4.2 GeV. 
}
\label{fig:ALLphipipq3}
\end{figure}

\begin{figure}[hbt]
\centerline{\includegraphics[height=7.0in,angle=90]{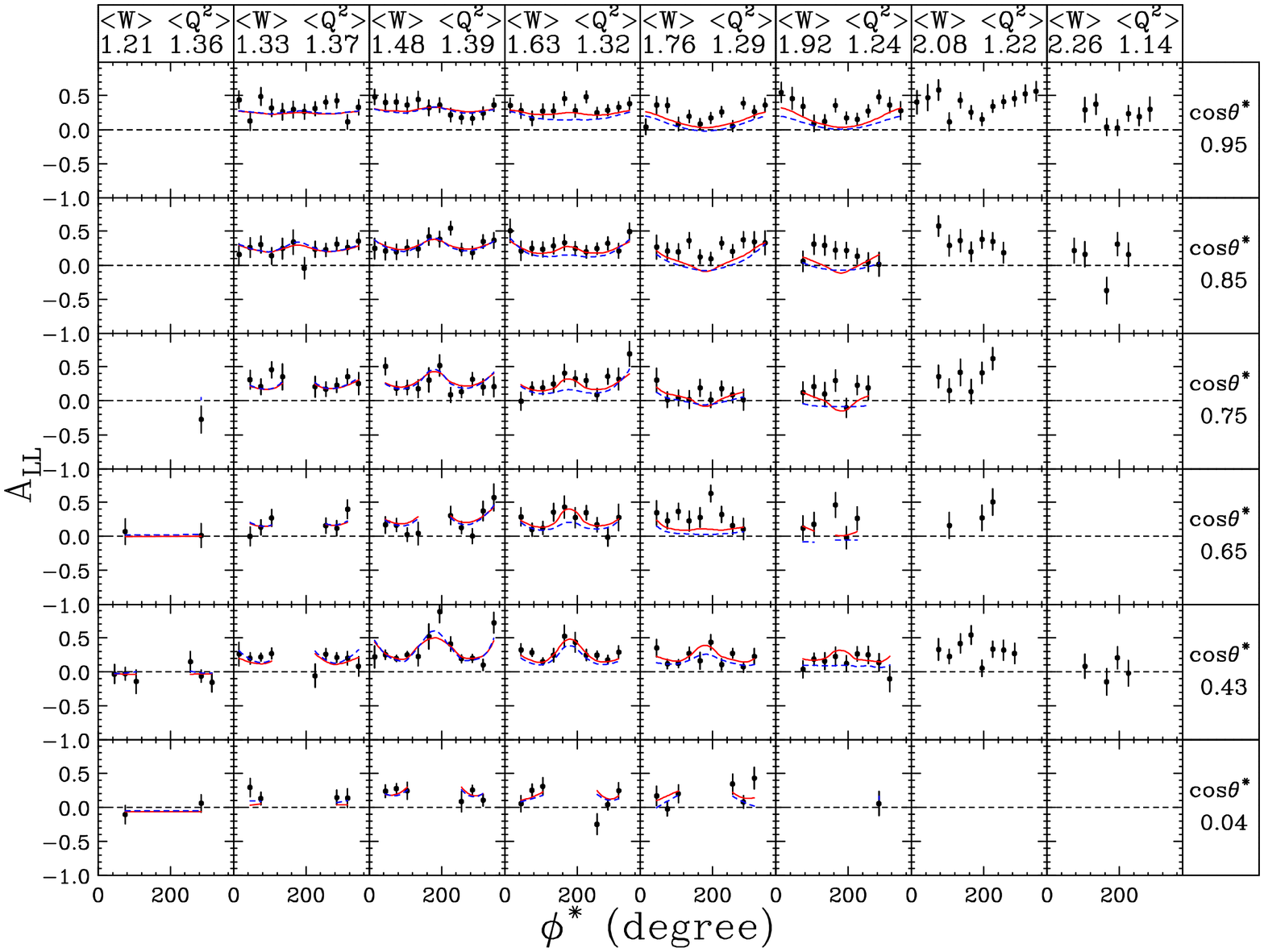}}
\caption{
Same as Fig.~\ref{fig:ALLphipipq1}, except for a beam energy
range of 5.7 GeV. 
}
\label{fig:ALLphipipq4}
\end{figure}

\subsection{$A_{UL}$ for \tpIn}

The results for $A_{UL}$ for \tpI are shown in 
Fig.~\ref{fig:AULphipipq1} (1.7 GeV beam energy),
Fig.~\ref{fig:AULphipipq2} (2.5 GeV),
Fig.~\ref{fig:AULphipipq3} (4.2 GeV), and 
Fig.~\ref{fig:AULphipipq4} (5.7 GeV).
Also shown on the plots are the
MAID 2007~\cite{maid} and JANR~\cite{janr} fits, averaged with
the same weighting as the data points.
By and large, MAID 2007 describes the data very well for
the 1.7 GeV beam energy data, including the dramatic
increase in \phicmsp dependence seen starting at $W=1.5$ GeV.
The magnitude of the  \phicmsp dependence is somewhat
underestimated at forward angles, however.
The JANR fit also describes the data well for $W<1.5$ GeV,
with increasingly large discrepancies at higher values of $W$.

For the higher beam energies, both fits are in reasonable
agreement with data only for $W<1.5$ GeV. At higher values
of $W$, disagreements generally become larger with increasing
beam energy (corresponding to higher values of $Q^2$.
In particular, the very large values 
of $A_{UL}$ observed for $1.7<W<2$ GeV and
$\cos(\theta^*)<0.8$ are not described by either fit to
previous data.
 
\begin{figure}[hbt]
\centerline{\includegraphics[height=7.0in,angle=90]{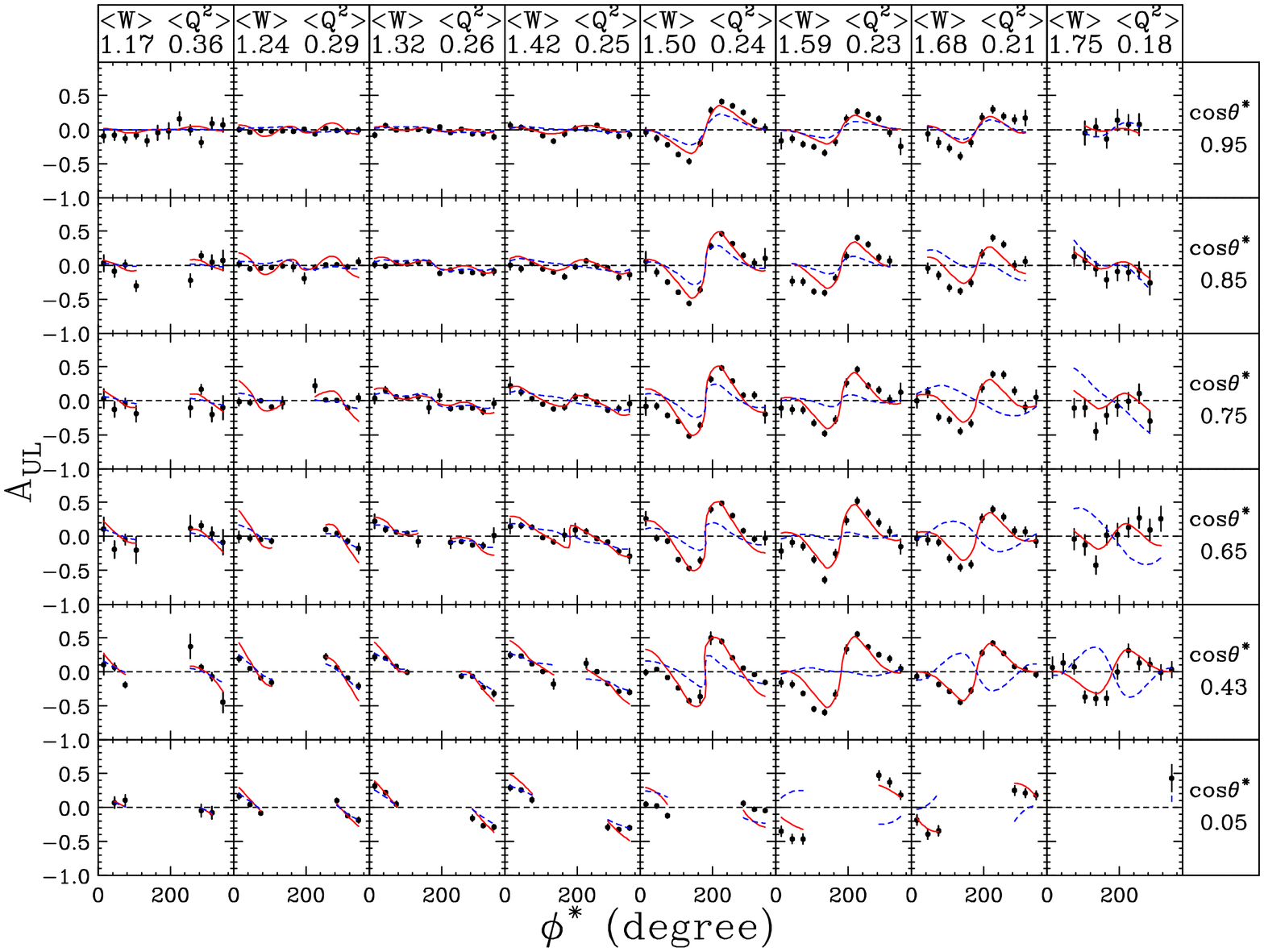}}
\caption{(color online) 
Results for $A_{UL}$ averaged over $Q^2$ as a function of $\phi^*$ 
in eight regions of $W$ (left to right) and the six
regions in \cthcmsp (top to bottom)
for the reaction \tpI and a beam energy range of 1.6 to 1.7 GeV.
The error bars do not include systematic uncertainties. 
The solid red curves are from MAID 2007 and the dashed blue
curves are from JANR~\cite{janr}.
Only results with uncertainties less than 0.2 are plotted, along with
the corresponding model curves. This results in some empty panels. 
}
\label{fig:AULphipipq1}
\end{figure}

\begin{figure}[hbt]
\centerline{\includegraphics[height=7.0in,angle=90]{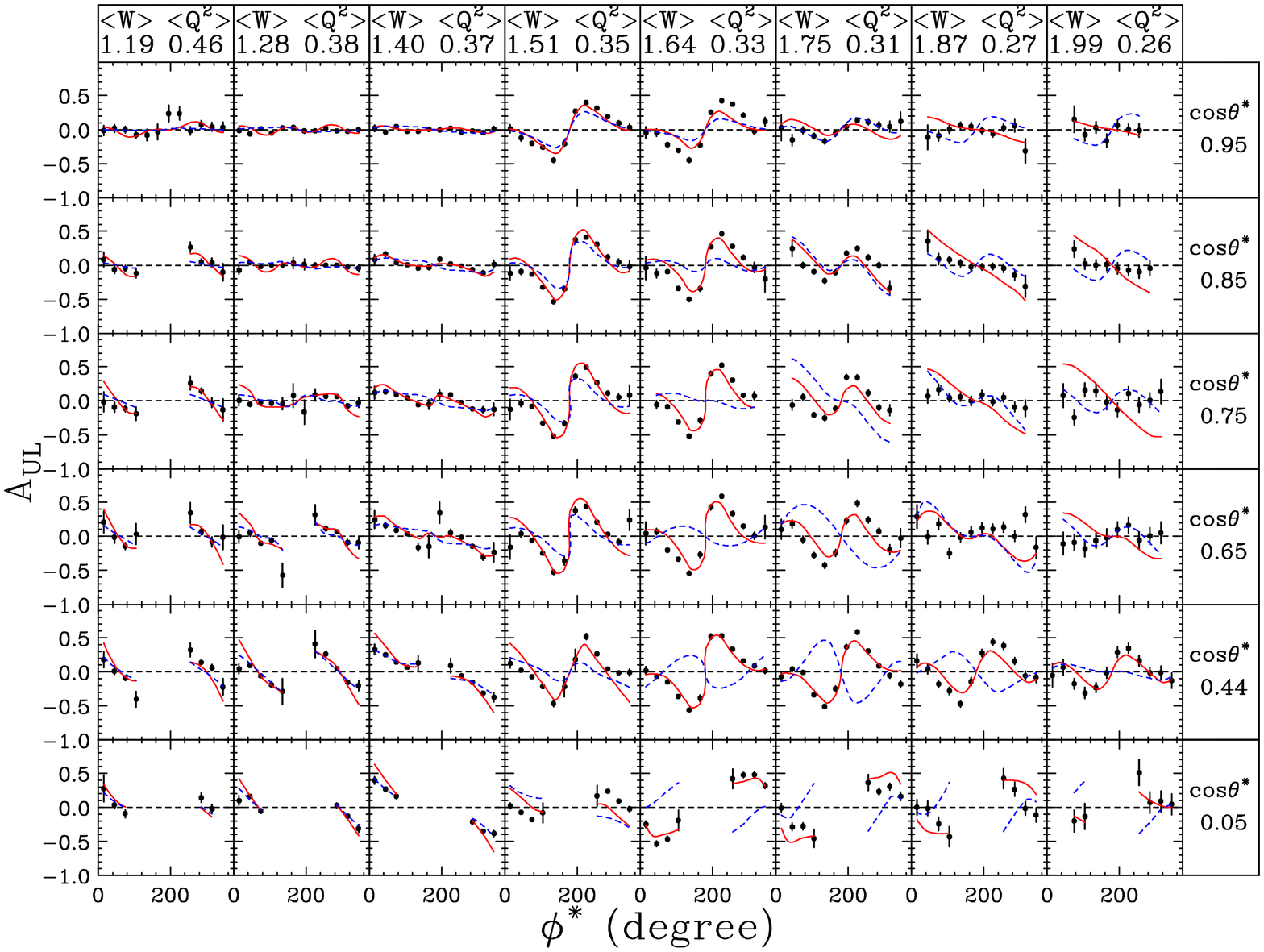}}
\caption{
Same as Fig.~\ref{fig:AULphipipq1}, except for a beam energy
range of 2.2 to 2.5 GeV. 
}
\label{fig:AULphipipq2}
\end{figure}

\begin{figure}[hbt]
\centerline{\includegraphics[height=7.0in,angle=90]{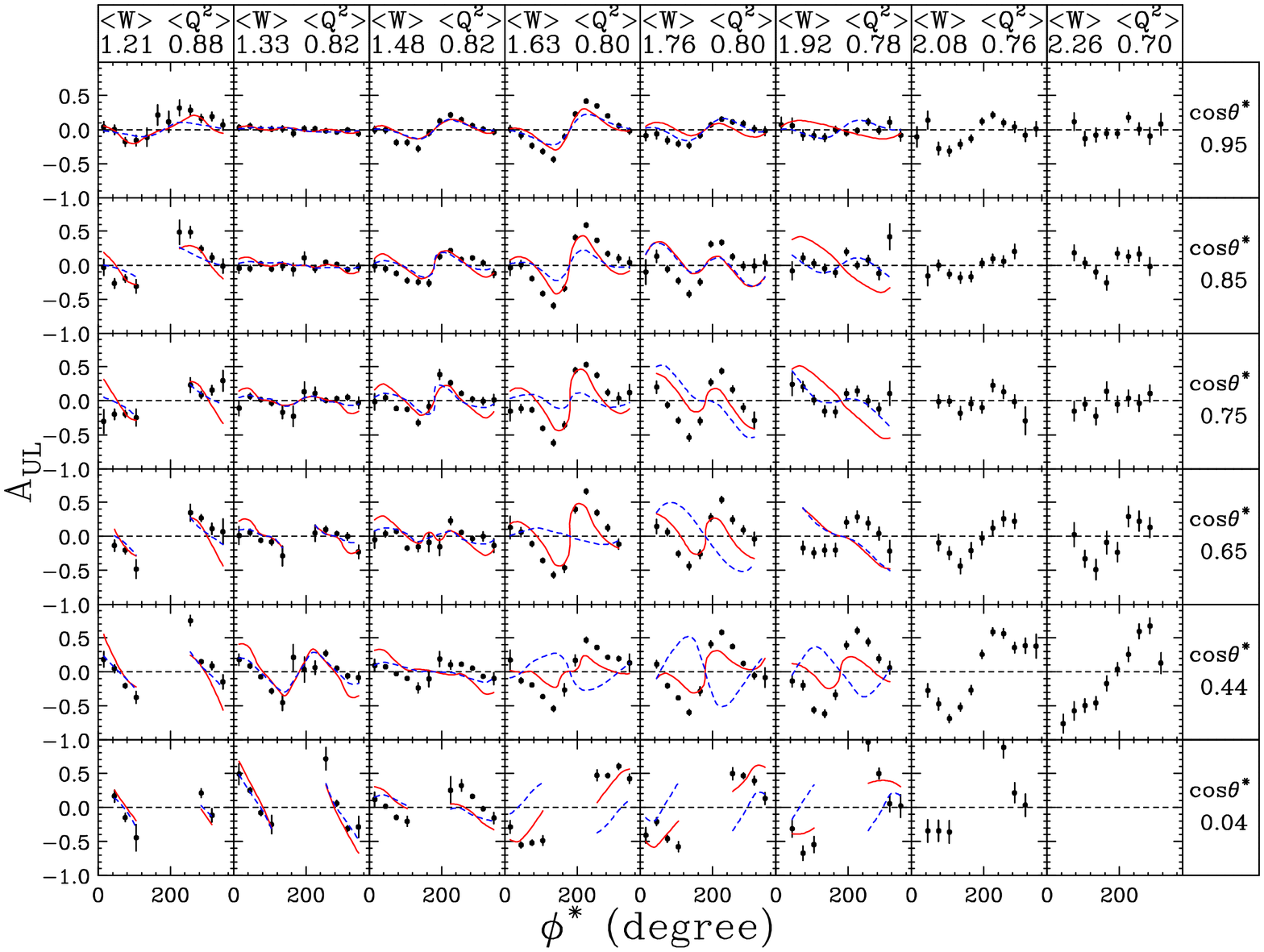}}
\caption{
Same as Fig.~\ref{fig:AULphipipq1}, except for a beam energy
range of 4.2 GeV. 
}
\label{fig:AULphipipq3}
\end{figure}

\begin{figure}[hbt]
\centerline{\includegraphics[height=7.0in,angle=90]{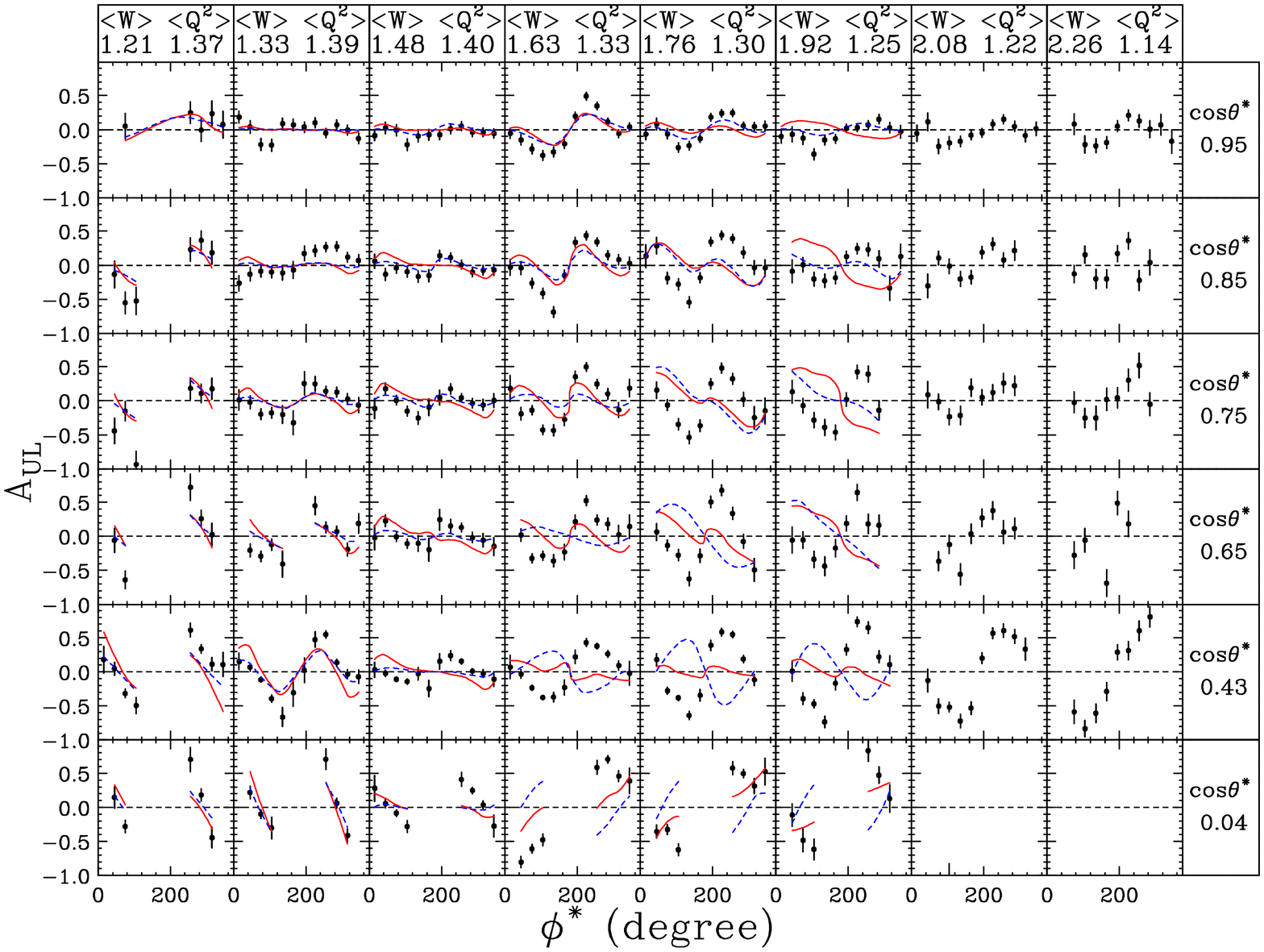}}
\caption{
Same as Fig.~\ref{fig:AULphipipq1}, except for a beam energy
range of 5.7 GeV. 
}
\label{fig:AULphipipq4}
\end{figure}

\subsection{$A_{LL}$ for \tpIVn}

The results for $A_{LL}$ for \tpIV are shown in 
Fig.~\ref{fig:ALLphipimq1} (1.7 GeV beam energy),
Fig.~\ref{fig:ALLphipimq2} (2.5 GeV),
Fig.~\ref{fig:ALLphipimq3} (4.2 GeV), and 
Fig.~\ref{fig:ALLphipimq4} (5.7 GeV).
Also shown on the plots are the
results of the MAID 2007 fit~\cite{maid}, averaged with
the same weighting as the data points. No final state
corrections have been applied to the model, nor has the
D-state component of the deuteron wave function been taken
into account in making this comparison. The JANR fit~\cite{janr}
is not available for this channel. 
By and large, MAID 2007 describes the data moderately well
although the model tends to be 
more negative than the data in bins where there is 
a difference. The largest discrepancy is for $W>1.7$ GeV
at forward angles and high $Q^2$, where a large
difference in the \phicmsp dependence can be seen. 
 
\begin{figure}[hbt]
\centerline{\includegraphics[height=7.0in,angle=90]{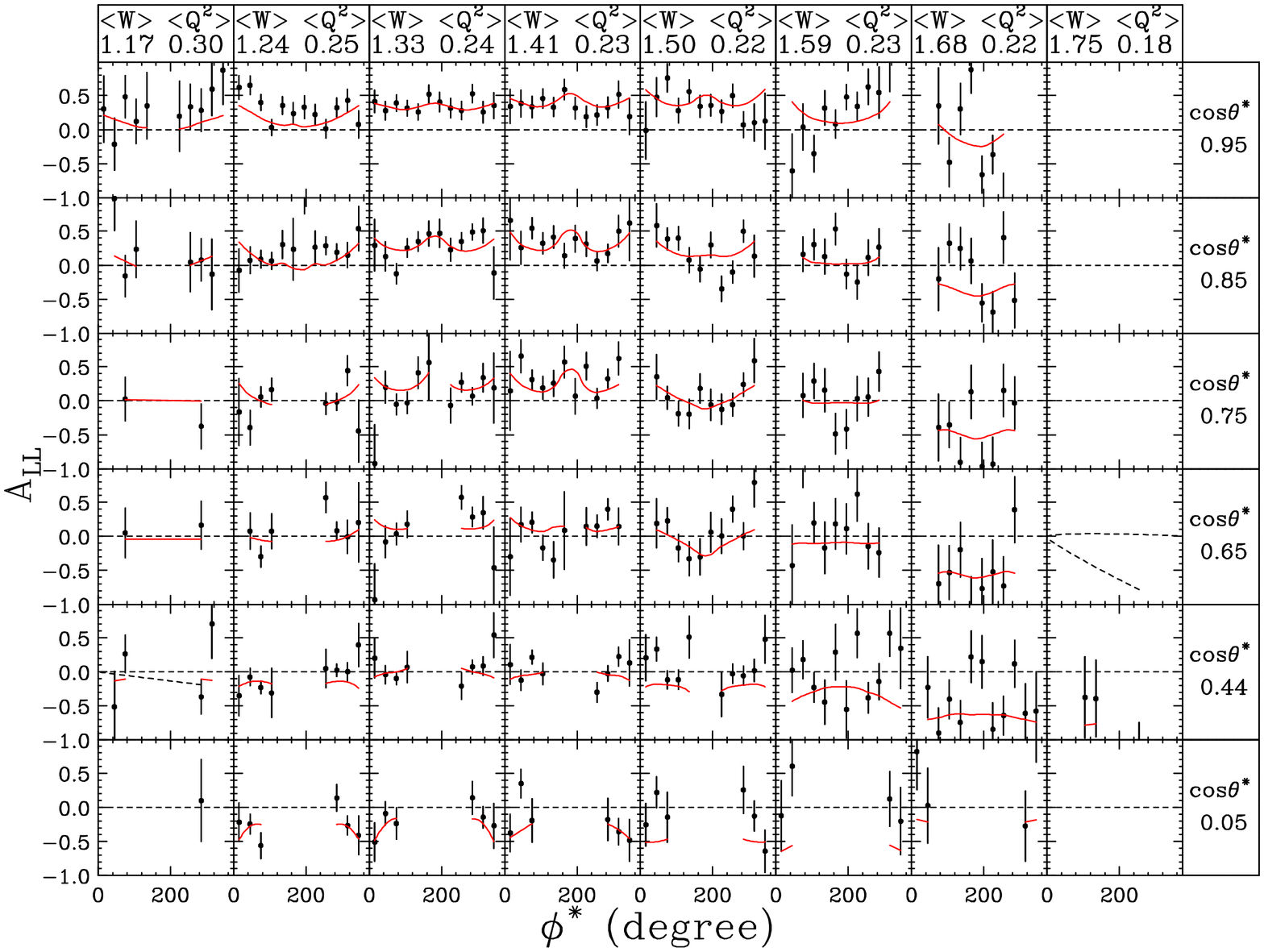}}
\caption{(color online) 
Results for $A_{LL}$ averaged over $Q^2$ as a function of $\phi^*$ 
in eight regions of $W$ (left to right) and the six
regions in \cthcmsp (top to bottom)
for the reaction \tpIV and a beam energy range of 1.6 to 1.7 GeV.
The error bars do not include systematic uncertainties. 
The curves are from MAID 2007.
Only results with uncertainties less than 0.6 are plotted, along with
the corresponding model curves. This results in some empty panels. 
}
\label{fig:ALLphipimq1}
\end{figure}

\begin{figure}[hbt]
\centerline{\includegraphics[height=7.0in,angle=90]{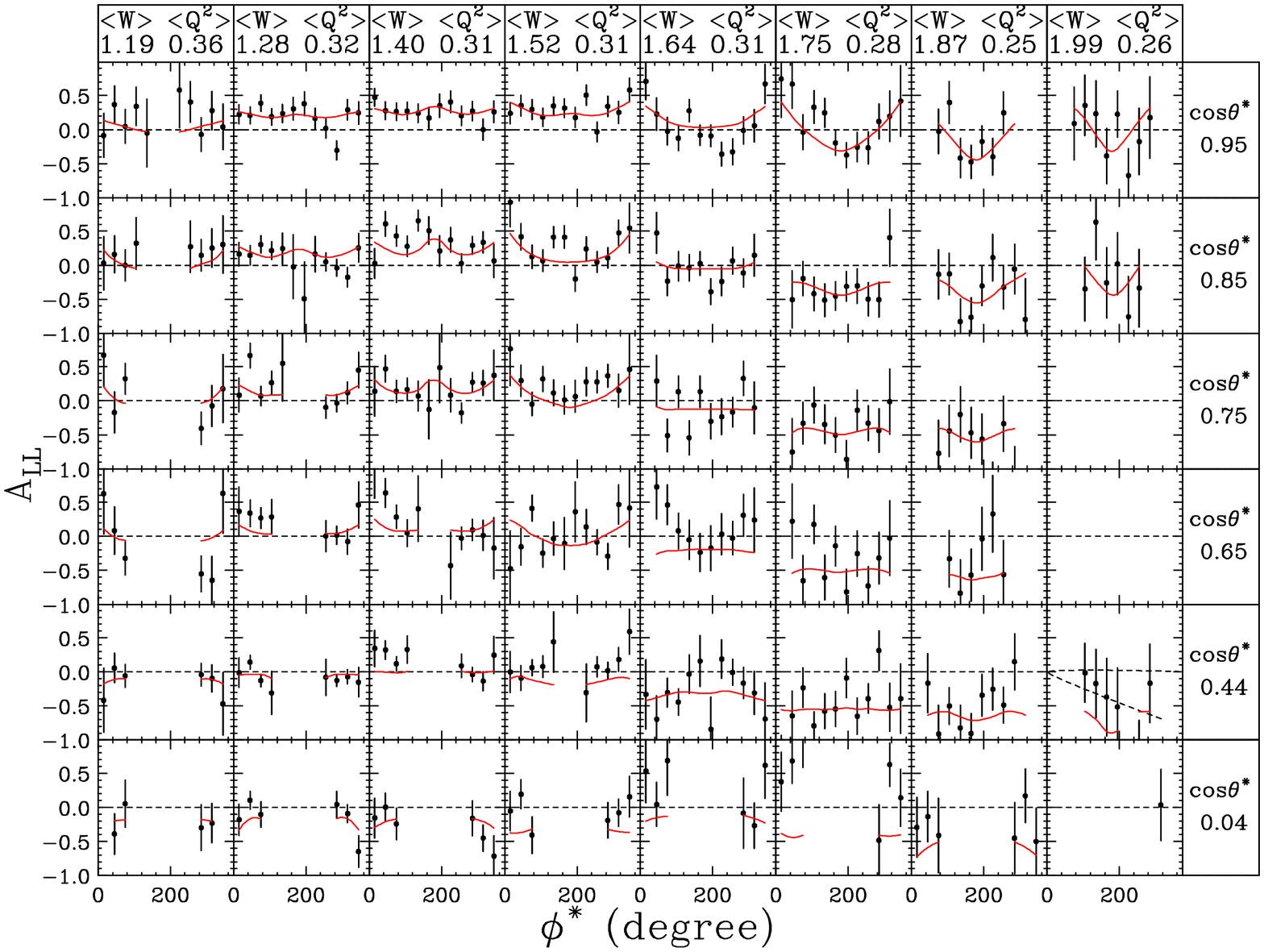}}
\caption{
Same as Fig.~\ref{fig:ALLphipimq1}, except for a beam energy
range of 2.2 to 2.5 GeV. 
}
\label{fig:ALLphipimq2}
\end{figure}

\begin{figure}[hbt]
\centerline{\includegraphics[height=7.0in,angle=90]{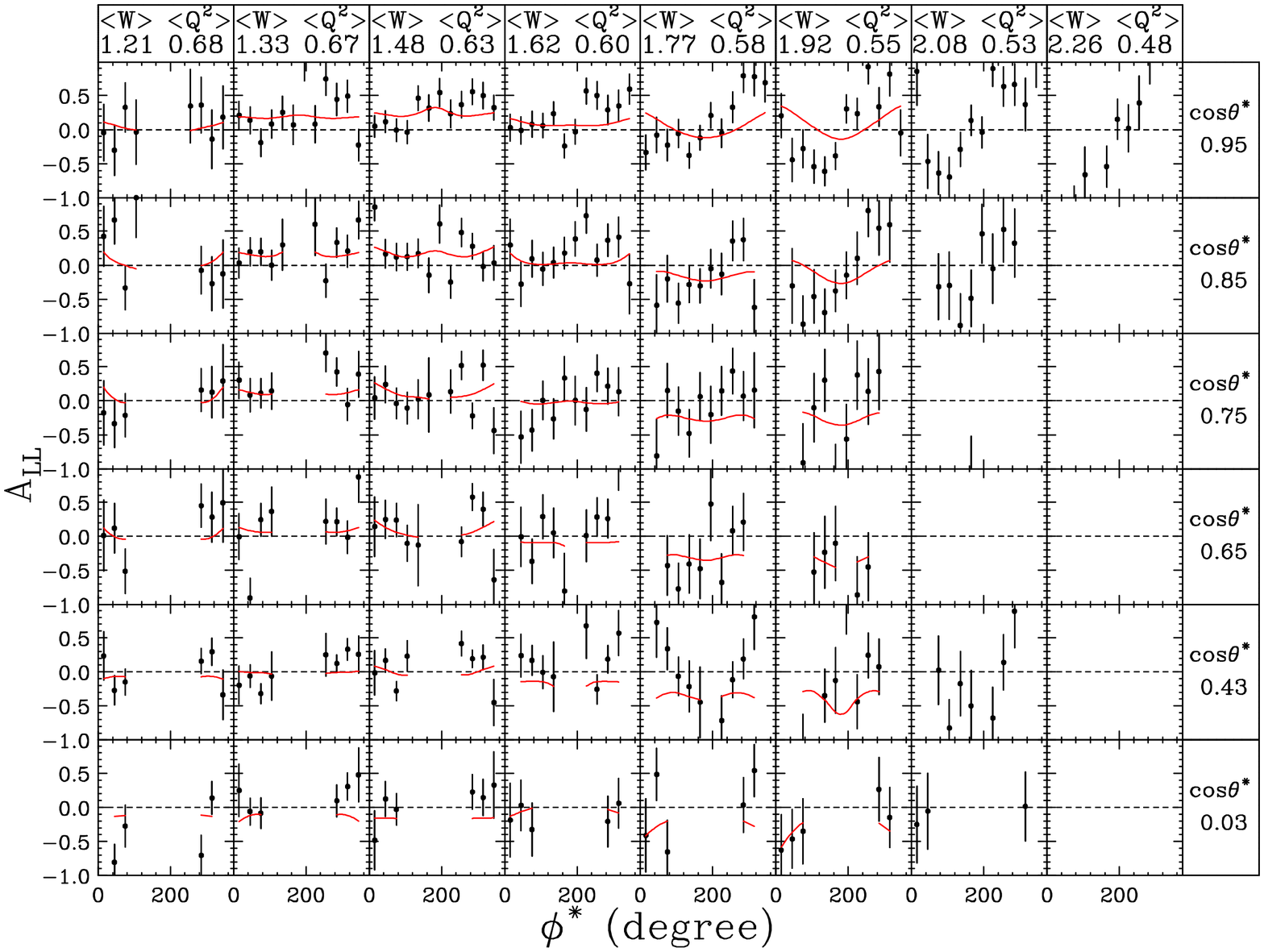}}
\caption{
Same as Fig.~\ref{fig:ALLphipimq1}, except for a beam energy
range of 4.2 GeV. 
}
\label{fig:ALLphipimq3}
\end{figure}

\begin{figure}[hbt]
\centerline{\includegraphics[height=7.0in,angle=90]{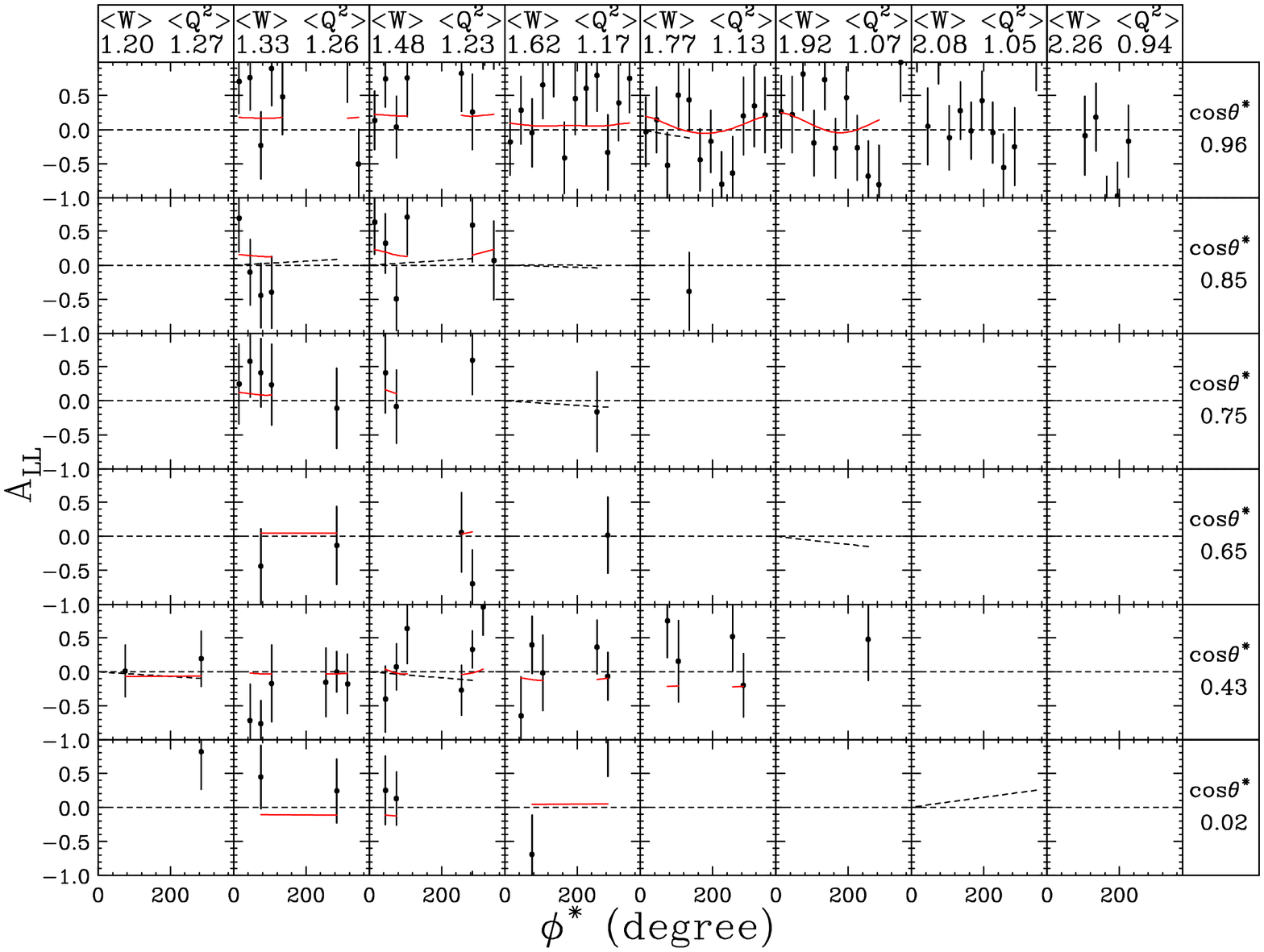}}
\caption{
Same as Fig.~\ref{fig:ALLphipimq1}, except for a beam energy
range of 5.7 GeV. 
}
\label{fig:ALLphipimq4}
\end{figure}

\subsection{$A_{UL}$ for \tpIVn}

The results for $A_{UL}$ for \tpIV are shown in 
Fig.~\ref{fig:AULphipimq1} (1.7 GeV beam energy),
Fig.~\ref{fig:AULphipimq2} (2.5 GeV),
Fig.~\ref{fig:AULphipimq3} (4.2 GeV), and 
Fig.~\ref{fig:AULphipimq4} (5.7 GeV).
Also shown on the plots are the
results of the MAID 2007 fit~\cite{maid}, averaged with
the same weighting as the data points. In this case,
the MAID fit sometimes describes the data moderately
well, but in most cases where a strong \phicmsp 
dependence is seen in the data, it is weaker in MAID
than in the data. This is particularly clear at
forward angles for $1.4<W<1.6$ GeV: a region that
is well described by MAID for the \tpI reaction,
but not the present \tpIV reaction. 
 
\begin{figure}[hbt]
\centerline{\includegraphics[height=7.0in,angle=90]{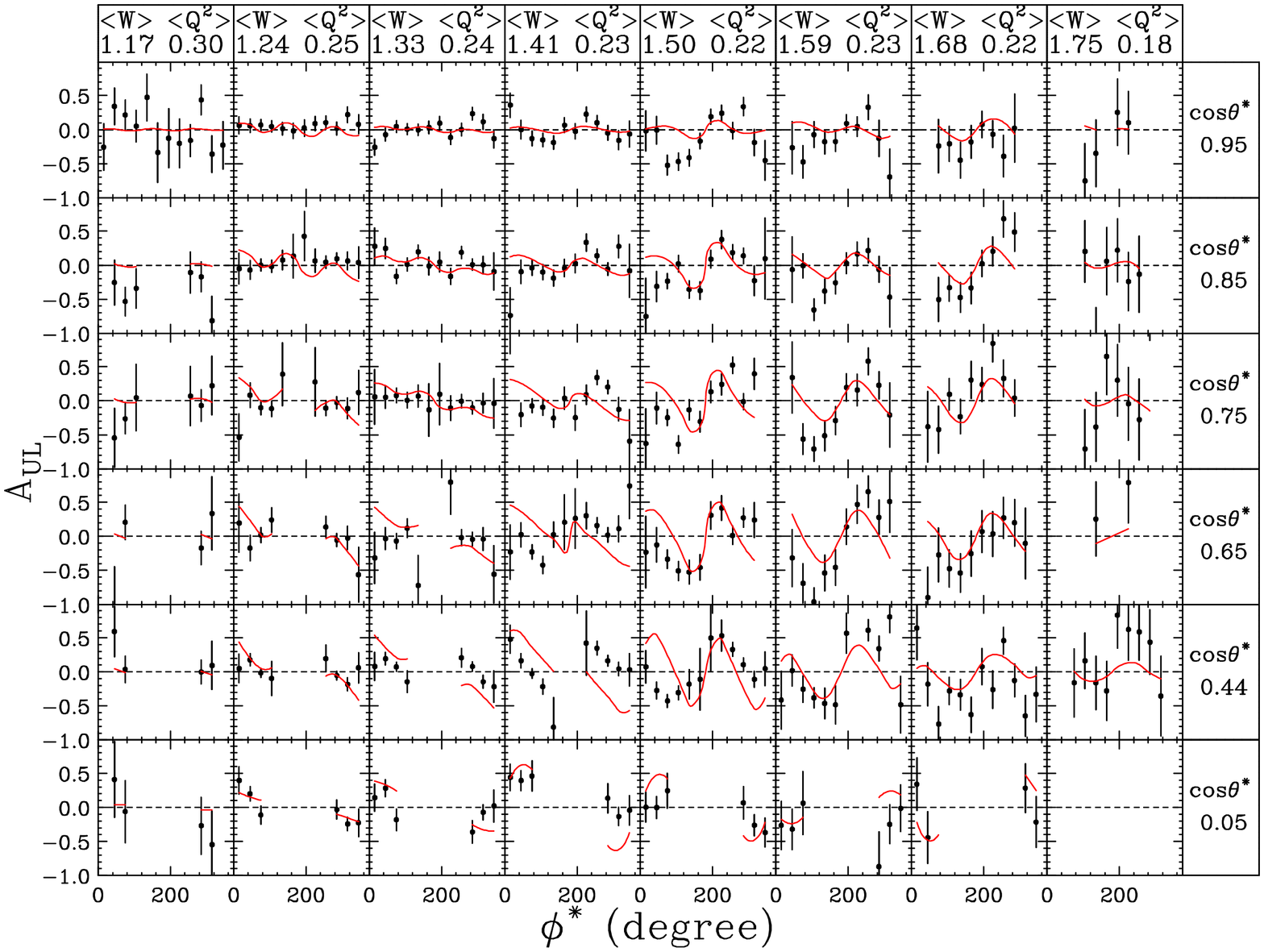}}
\caption{(color online) 
Results for $A_{UL}$ averaged over $Q^2$ as a function of $\phi^*$ 
in eight regions of $W$ (left to right) and the six
regions in \cthcmsp (top to bottom)
for the reaction \tpIV and a beam energy range of 1.6 to 1.7 GeV.
The error bars do not include systematic uncertainties. 
The curves are from MAID 2007.
Only results with uncertainties less than 0.6 are plotted, along with
the corresponding model curves. This results in some empty panels. 
}
\label{fig:AULphipimq1}
\end{figure}

\begin{figure}[hbt]
\centerline{\includegraphics[height=7.0in,angle=90]{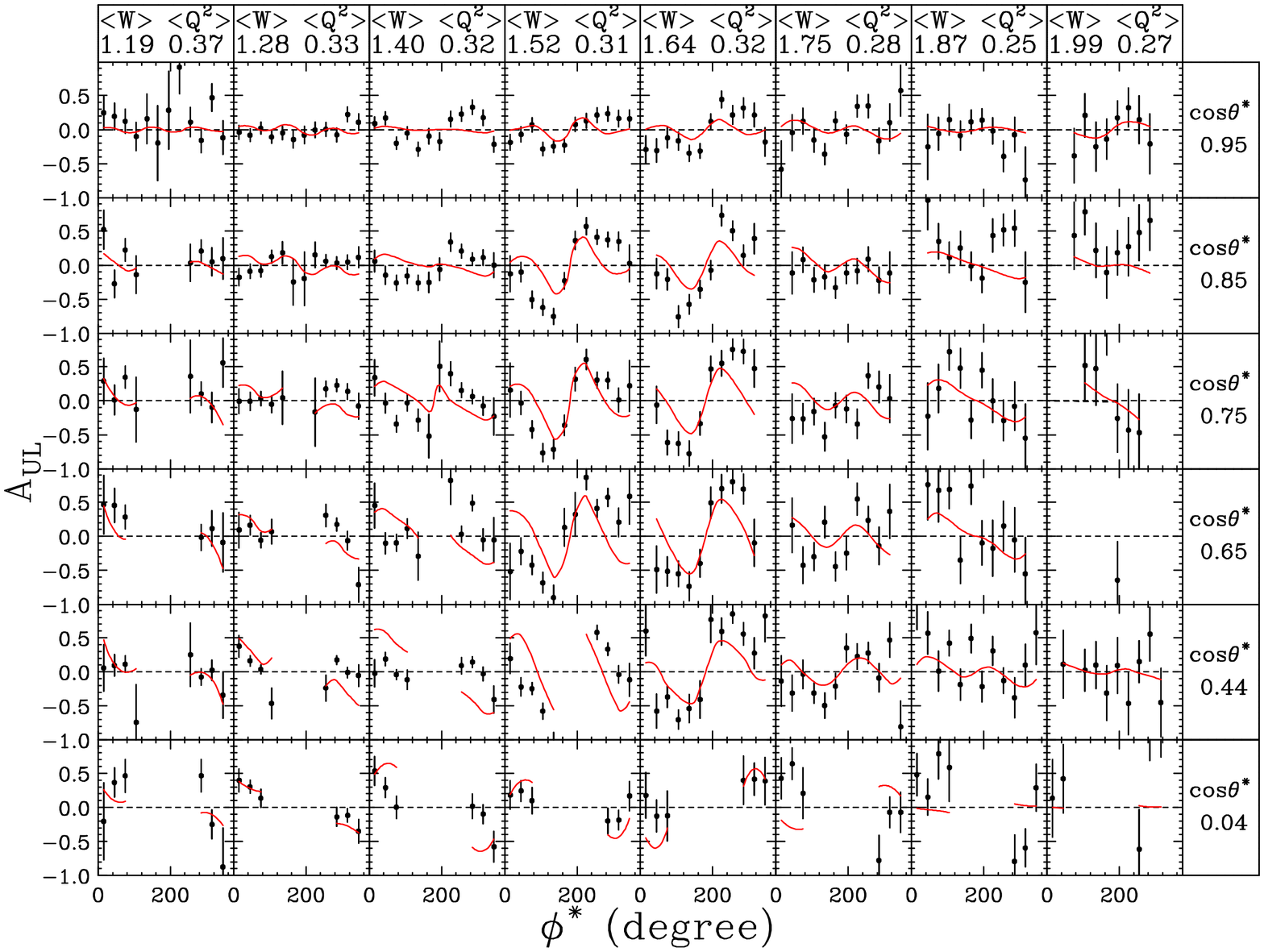}}
\caption{
Same as Fig.~\ref{fig:AULphipimq1}, except for a beam energy
range of 2.2 to 2.5 GeV. 
}
\label{fig:AULphipimq2}
\end{figure}

\begin{figure}[hbt]
\centerline{\includegraphics[height=7.0in,angle=90]{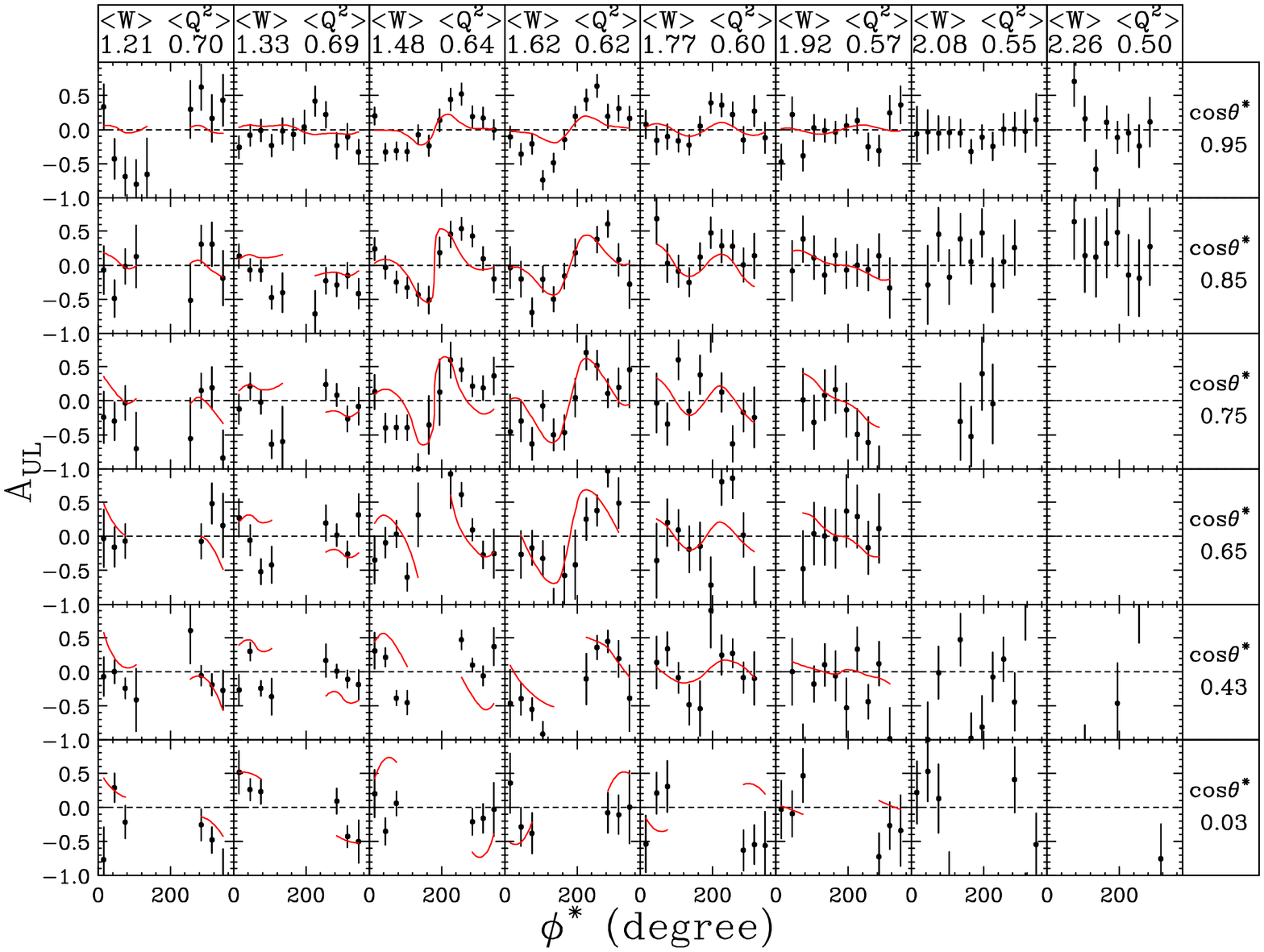}}
\caption{
Same as Fig.~\ref{fig:AULphipimq1}, except for a beam energy
range of 4.2 GeV. 
}
\label{fig:AULphipimq3}
\end{figure}

\begin{figure}[hbt]
\centerline{\includegraphics[height=7.0in,angle=90]{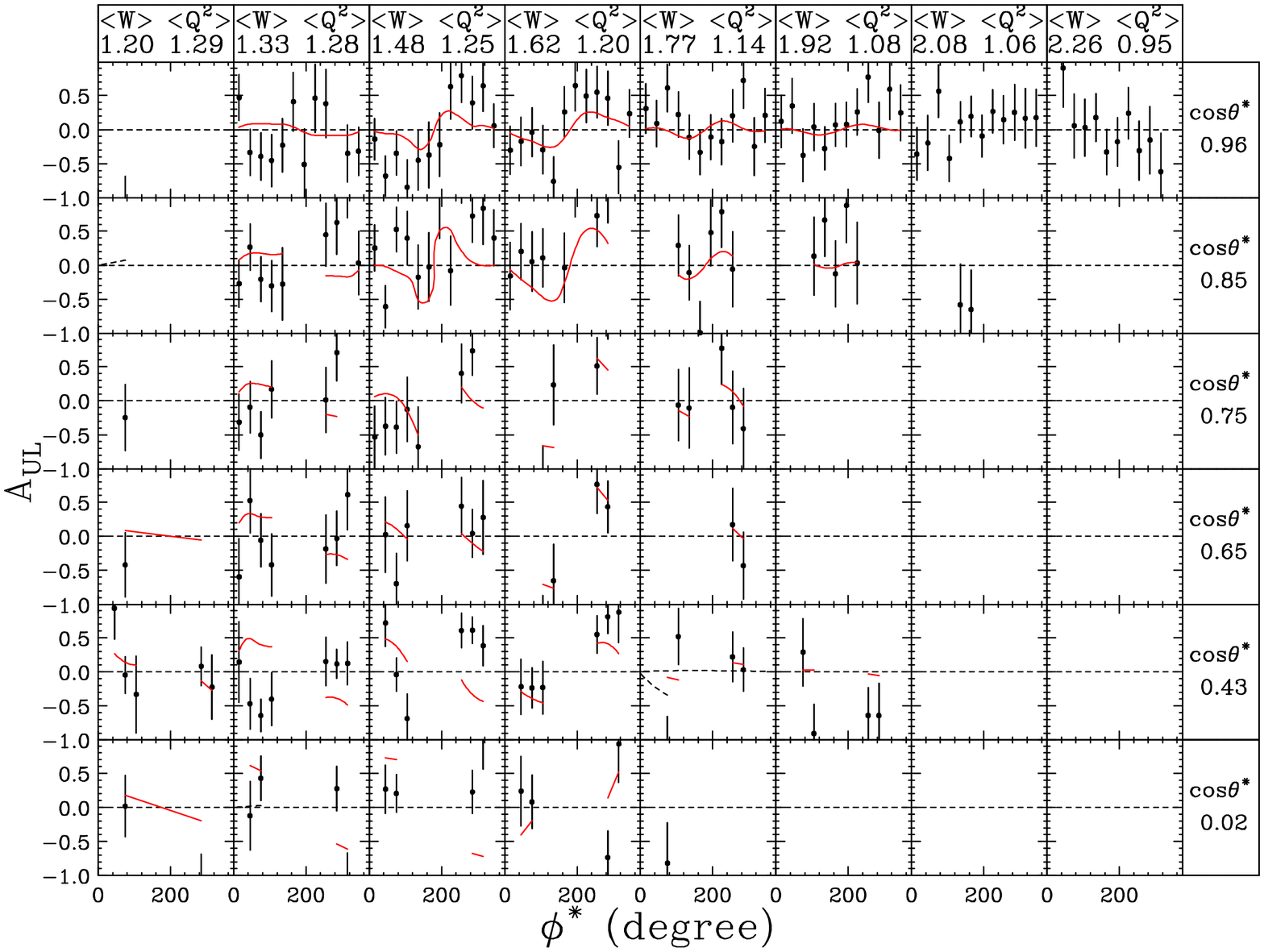}}
\caption{
Same as Fig.~\ref{fig:AULphipimq1}, except for a beam energy
range of 5.7 GeV. 
}
\label{fig:AULphipimq4}
\end{figure}

%bsa ALU moved to exclnoteALU.tex

\section{Summary}
Beam-target double spin asymmetries and target single-spin
asymmetries in exclusive  $\pi^+$ and $\pi^-$  electroproduction
were obtained from scattering of 1.6, 1.7, 2.2, 2.5,
4.2, and 5.7 GeV longitudinally polarized
electrons from longitudinally polarized protons and deuterons
using the CLAS detector at Jefferson Lab. 
The kinematic range covered is $1.1<W<2.6$ GeV and $0.05<Q^2<5$
GeV$^2$, greatly expanding the range of previous data.
The asymmetry results are 
presented in large data tables which are suitable, for
example, as input to the calculations of
radiative corrections to semi-inclusive pion electroproduction.
When used to make improved empirical fits, 
the data will provide powerful constraints on 
the $Q^2$-dependence  of $N^*$ and $\Delta^*$ resonance
amplitudes and phases,
and the interplay with non-resonant contributions.
The higher $W$ coverage compared with previous data
may reveal the importance of previously poorly-described
nucleon resonances. 

In comparison with the MAID 2007 and JANR fits, we find good
agreement for the \tpI asymmetries for $W<1.7$ GeV
and $Q^2<1$ GeV$^2$, a kinematic region where many
data were available as input to this fit. For $W>1.7$
GeV and higher values of $Q^2$, some large discrepancies
with MAID are observed, particularly in the target-spin
asymmetry.  In the case of the \tpIV reaction, 
significant discrepancies with MAID are seen at all
values of $W$, especially in $A_{UL}$, which is not
too surprising as very few data were available prior
to the present experiment to constrain fits such as
MAID 2007. Clearly the new data presented in this analysis
will provide powerful new constraints on global
fits. 

\section*{Acknowledgments}
We thank Inna Aznauryan for providing the JANR source
code and L. Tiator for providing the MAID 2007
source code. 
%We thank X. Zheng for suggesting the functional form of
%the dilution factor fit, and for careful review of the analysis.
We acknowledge the outstanding efforts of the staff
of the Accelerator and the Physics Divisions at Jefferson Lab that made
this experiment possible.  This material is based
upon work supported by the U.S. Department of Energy, 
Office of Science, Office of Nuclear Physics under contract 
DE-AC05-06OR23177 and the National Science Foundation. Partial
support was provided by 
the Scottish Universities Physics Alliance (SUPA),
the United Kingdom's Science and Technology Facilities Council,
the National Research Foundation of Korea,
the Italian Instituto Nazionale di Fisica Nucleare, the French Centre
National de la Recherche Scientifique, the French Commissariat \`{a}
l'Energie Atomique,  and 
the Emmy Noether grant from the Deutsche Forschungsgemeinschaft.
The Southeastern Universities Research Association (SURA) operates
the Thomas Jefferson National Accelerator Facility for the
United States Department of Energy under contract DE-AC05-84ER-40150.


\begin{thebibliography}{9}
\bibitem{bebek76} C. J. Bebek {\em et al.}, 
Phys. Rev. D {\bf 13}, 25  (1976).

\bibitem{bebek78} C. J. Bebek {\em et al.}, 
Phys. Rev. D {\bf 17}, 1693  (1978).

\bibitem{Horn09} T. Horn {\em et al.}, 
Phys. Rev. C \textbf{78}, 058201 (2008).

\bibitem{HPBlok} H. P. Blok {\em et al.}, 
   Phys. Rev. C \textbf{78}, 045202 (2008).

\bibitem{XQian} X. Qian {\em et al.}, 
Phys. Rev. C \textbf{81}, 055209 (2010).

\bibitem{ParkA} H. Egiyan {\em et al.} [CLAS Collaboration],
   Phys. Rev. C \textbf{73}, 025204 (2006);
K. Park {\it et al.,} [CLAS Collaboration], Phys. 
Rev. C {\bf 77}, 015208 (2008).

\bibitem{ParkB} K. Park {\em et al.} [CLAS Collaboration], 
   Phys. Rev. C \textbf{91}, 045203 (2015).

\bibitem{ParkC} K. Park {\em et al.} [CLAS Collaboration], 
    Eur. Phys. J. {\bf A49}, 16 (2013).

\bibitem{morris}
J. Morris {\it et al.,} Phys. Lett. \textbf{B73}, 495 (1978);
J. Morris {\it et al.,} Phys. Lett. \textbf{B86}, 211 (1979).

\bibitem{Fosteretc}
F. Foster and G. Hughes, Rep. Prog. Phys. \textbf{46}, 1445 (1983);
O.~Varenikova {\it et al.,} Z. Phys. C \textbf{37}, 251 (1988);
J.~Wright {\it et al.,} Nucl. Phys. \textbf{B181}, 403 (1981);
D.~Gaskell {\it et al.,} Phys. Rev. Lett. \textbf{87}, 202301 (2001).

%\bibitem{Hermes} A. Airapetian {\em et al.}, 
%Phys. Lett. B \textbf{659}, 486 (2008).

\bibitem{Avakian} 
H. Avakian and L. Elouadrhiri, Proc. High Energy Spin
Physics, A.V. Efremov and O.V. Teryaev Editors, Dubna
(2003).

\bibitem{daVita}
R. De Vita {\it et al.} 
[CLAS Collaboration], Phys. Rev. Lett. \textbf{88}, 082001 (2002). 
 
\bibitem{eg4}
X. Zheng {\it et al.}  [CLAS Collaboration], 
arXiv:1607.03924 (2016), 
accepted in Phys. Rev. C. 

\bibitem{eg1dvcs}
P. Bosted {\it et al.}  [CLAS Collaboration], 
arXiv:1604.04350 (2016),
submitted to Phys. Rev. C.


%\bibitem{recoilpol}
%Th. Pospischil et al., Phys. Rev. Lett. \textbf{86}, 2959 (2001).

\bibitem{maid} 
www.portal.kph.uni-mainz.de/MAID/;
D. Drechsel,  O. Hanstein, S.S. Kamalov, L. Tiator, 
Nucl. Phys. \textbf{A645}, 145 (1999). 

\bibitem{Keith} C.D. Keith {\it et al.,} Nucl. Instr. Meth. 
{\bf 501}, 327 (2003). 
 
\bibitem{CLAS} B.A. Mecking {\it et al.}, 
Nucl. Instr. Meth., {\bf 503}, 513 (2003).

\bibitem{prok}
  Y.~Prok {\it et al.}  [CLAS Collaboration],
  Phys.\ Lett. \ {\bf B672}, 12 (2009).

\bibitem{vipuli}
  K.~V.~Dharmawardane {\it et al.}  [CLAS Collaboration],
  Phys. Lett. \textbf{B641}, 11 (2006).

\bibitem{duality}
P.E. Bosted  {\it et al.} [CLAS Collaboration],
Phys. Rev. \ C \textbf{75}, 035203 (2007). 

\bibitem{inclp}
R.G. Fersch, {\it et al.}  [CLAS Collaboration],
to be published; 
R.G. Fersch,, Ph.D thesis, College of William and Mary (2008),
www.Jlab.org/Hall-B/general/thesis/Fersch\_thesis.pdf

\bibitem{incld}
N. Guler {\it et al.}  [CLAS Collaboration] , 
Phys. Rev. C {\bf 92}, 055201 (2015)
[arXiv:1505.07877]

\bibitem{pierce}
J. Pierce, Ph.D thesis, University of Virginia, 2008 (unpublished).

\bibitem{sharon}
S. Careccia, Ph.D thesis, Old Dominion University, 2012 (unpublished).

\bibitem{bissellieg1b}
A. Biselli {\it et al.}  [CLAS Collaboration],
Phys. Rev. C \textbf{78}, 045204 (2008).

\bibitem{clasdb}
CLAS data base: clasweb.jlab.org/physicsdb (search for
experiments eg1bexclpip and eg1bexclpim).

\bibitem{SMp}
See Supplemental Material at [URL will be inserted by publisher]
for plain text tables of asymmetry results for the reaction
\tpIn.

\bibitem{SMm}
See Supplemental Material at [URL will be inserted by publisher]
for plain text tables of asymmetry results for the reaction
\tpIVn.

\bibitem{Adep}
P.E.~Bosted and V.~Mamyan, [arXiv:1203.2262] (2012); 
P.E.~Bosted {\it et al.}  [CLAS Collaboration], 
Phys.\ Rev.\ C {\bf 78}, 015202 (2008).

\bibitem{tsai}
Y.~S.~Tsai, Report No.\ SLAC--PUB--848 (1971);
Y.~S.~Tsai, Rev.\ Mod.\ Phys.\ {\bf 46}, 815 (1974).

\bibitem{E143}
K. Abe {\it et al.}  [SLAC E143 Collaboration], 
Phys. \ Rev.  D {\bf 58},  112003 (1998).

%\bibitem{n15}
%P.~E.~Bosted {\it et al.}  [CLAS Collaboration],
%Phys.\ Rev.\ C {\bf 78}, 015202 (2008).

%\bibitem{arxiv} 
%P.~E.~Bosted {\it et al.}  [CLAS Collaboration],
%arXiv:1604.04350 [nucl-ex], files exclpip.txt and exclpim.txt.

\bibitem{janr}
I. G. Aznauryan, Phys. Rev. C \textbf{67}, 015209 (2003);
I.~G.~Aznauryan {\it et al.}  [CLAS Collaboration], 
Phys. Rev. C \textbf{80}, 055203 (2009).



%\bibitem{biselli}
%A. Biselli {\it et al.}  [CLAS Collaboration] ,
%Phys. Rev.C68, 035202 (2003).

%\bibitem{Oldsen}
% H. Oldsen, Phys. Rev. C  {\bf 144}, 887 (1959).

%\bibitem{DeMasi}
%R. De Masi {\it et al.}, (CLAS Collaboration), 
%Phys. Rev. C {\bf 77} (2008) 042201.
%[arXiv:0711.4736]

%\bibitem{Hermespip}
%A. Airapetian {\it et al.,}, (HERMES Collaboration),
%Phys. Lett. B535 (2002) 85.

%\bibitem{e1dvcseta}
%Bo Zhao {\it et al.} (CLAS Collaboration), to be published.

\end{thebibliography}
\end{document}